\pdfoutput=1

\RequirePackage{fix-cm}
\documentclass[twocolumn,epjc3]{svjour3}
\smartqed
\usepackage{graphicx}
\usepackage{amsmath,amssymb}
\usepackage{booktabs}
\usepackage{enumitem}
\usepackage{cite}
\usepackage{microtype}

\journalname{Eur. Phys. J. C}

\newcommand{\bn}{\mathbf{n}}
\newcommand{\bnp}{\mathbf{n}_{\perp}}
\newcommand{\bw}{\mathbf{w}}
\newcommand{\by}{\mathbf{y}}
\newcommand{\lmin}{\lambda_{\perp}^{(0)}}
\newcommand{\Vopt}{\mathcal{V}_{\rm opt}}

\begin{document}

\title{Log-periodic radiation cores and light rings in pure quadratic bulk gravity}

\author{Sudan Hansraj\thanksref{e1,addr1}
\and Rio Bissessur\thanksref{addr1}}

\thankstext{e1}{e-mail: hansrajs@ukzn.ac.za}

\institute{Astrophysics Research Centre, School of Mathematics, Statistics and Computer Science, University of KwaZulu-Natal, Private Bag X54001, Durban 4000, South Africa\label{addr1}}

\date{Received: date / Accepted: date}

\maketitle

\begin{abstract}
In the pure quadratic model of the B\"ohmer--Jensko bulk--boundary formulation of gravity, every static metric with a regular centre has vanishing central density and pressure, so no affine equation of state admits a regular non-vacuum star. We therefore study singular cores, in scale-invariant variables in which the isotropy condition is a single autonomous relation between two functions. Closures that keep the self-similar solutions as equilibria fix the stability of two vacuum lines but leave that of the matter-bearing equilibria undetermined, except at one unphysical point, so only an equation of state yields definite dynamics. For a linear equation of state the closed system has a focus, a saddle, and a vacuum point with nonzero bulk scalar that exchanges stability with the saddle in a transcritical bifurcation. At the radiation value the system is integrable, through the Tolman integral and scale invariance, and the focus becomes a nonlinear centre surrounded by log-periodic solutions. Radiation is also exactly where the line containing the matter-bearing equilibria of this closure coincides with the line of circular null orbits. Consequently every periodic radiation core with monotonic areal radius has exactly one stable and one unstable light ring per period, at radii forming two interlaced geometric progressions towards the centre with a common ratio tending to 2.165 for small oscillations. In the optical geometry the self-similar solution is a cylinder and each core a periodic tube, whose bulges and necks carry these rings. MIT bag-model configurations whose cores approach these orbits inherit this structure.
\end{abstract}

\section{Introduction}
\label{sec:intro}

Circular null orbits, or light rings, are a characteristic feature of ultracompact objects. In a static spherically symmetric spacetime they lie at the stationary points of the optical potential $e^{\nu}/R^2$, where $e^{\nu}=-g_{tt}$ and $R$ is the areal radius. Light rings of smooth horizonless ultracompact objects come in pairs, one of which is stable~\cite{Cunha2017}, and stable light rings have been associated with long-lived modes and possible nonlinear instabilities~\cite{Cardoso2014,Keir2016}. For a perfect fluid with $p=\gamma\rho$ the conservation equation gives $e^{\nu}\propto\rho^{-2\gamma/(1+\gamma)}$, so for a self-similar fluid, whose density is a power of $R$, the optical potential is a power of $R$ as well. In general relativity self-similar fluids have $\rho\propto R^{-2}$, and the optical potential is constant, with a circular null orbit at every radius, only for stiff matter, $\gamma=1$. Karlovini, Rosquist and Samuelsson used the Nilsson--Uggla formulation~\cite{Nilsson2001} to construct perfect-fluid configurations with arbitrarily many necks in the optical geometry~\cite{Abramowicz1988,Abramowicz1997}, each associated with an unstable circular null orbit~\cite{KRS2000,KRS2001}. In the pure quadratic model of bulk gravity studied here, self-similar fluids have $\rho\propto R^{-4}$, and the same argument selects radiation, $\gamma=\tfrac13$. We show that radiation is also the value at which the field equations closed by the equation of state become integrable, and that the periodic solutions surrounding the self-similar one resolve its degeneracy into infinitely many alternating stable and unstable light rings with exactly geometric spacing.

The model belongs to a broad effort to extend general relativity, motivated by the late-time accelerated cosmic expansion, by the limited tests of the theory at cosmological length scales and in extreme gravity regimes, and by its non-renormalisability~\cite{Goroff1985,tHooft1974}. Leading proposals include Lovelock gravity~\cite{love1,love2}, whose second-order member, Einstein--Gauss--Bonnet gravity, is free of Ostrogradsky instabilities but dynamical only in more than four dimensions, despite contested attempts to rescale its coupling to four dimensions~\cite{Glavan,GursesSismaniTekin}; $f(R)$ gravity~\cite{Starobinsky,Nojiri2007,Sotiriou2010,DeFelice2010}, which has fourth-order field equations and in which relativistic stars with prescribed equations of state have been constructed numerically~\cite{Babichev2010}; and $f(R,T)$~\cite{Harko2011}, Rastall~\cite{Rastall1972} and unimodular gravity~\cite{HansrajEllis1}, which allow non-conservation of energy--momentum, although unimodular gravity can equally be formulated with conserved matter~\cite{Bufalo2015}.

B\"ohmer and Jensko~\cite{BJ} proposed to unify classes of modified gravity theories through the role of the boundary term in the decomposition of the Ricci scalar into a bulk term $G$ and a boundary term $B$. Metric-affine theories admit formulations, such as the teleparallel equivalent of general relativity~\cite{Maluf2013}, whose action is Lorentz invariant only up to a boundary term, as is the Einstein--Hilbert action, and a theory built from the bulk term alone yields field equations of at most second order. Boundary terms were introduced in teleparallel gravity through the $f(T,B)$ models of~\cite{Bahamonde2015}, in which $f(R)$ is the unique Lorentz invariant case and $f(T)$ the unique second-order case. B\"ohmer and Jensko identified three types of boundary term that admit $f(T)$, $f(Q)$ and $f(R)$ gravity as limiting cases~\cite{BJ} and gave a unified treatment in the Palatini formalism~\cite{BJ2}, and boundary terms have since been studied in $f(Q,B)$ gravity~\cite{Capozziello2023}, with dynamical systems methods~\cite{Shabani2025,Kadam2023}, and in higher-curvature models~\cite{Vishwakarma2024}. To avoid confusion with $f(\mathcal{G})$ gravity, where $\mathcal{G}$ is the Gauss--Bonnet invariant, we refer to the model of~\cite{BJ} as $f(G,B)$ theory; since we take $f$ independent of $B$, we also call it bulk gravity.

Isotropic stars, in which the radial and tangential pressures coincide, must satisfy the pressure isotropy condition, which makes the field equations harder to solve; nevertheless many exact solutions are known in general relativity~\cite{finchskea,delgaty,Tolman1939,Buchdahl1959,Ivanov2002} and in Gauss--Bonnet models~\cite{N2Dadhich,Hansrajt1}. When exact solutions are out of reach, an autonomous form of the equations reveals the behaviour of all solutions through its equilibria and their stability~\cite{BanasiakArlotti,Perko}. For static spherically symmetric perfect fluids in general relativity, homology-invariant variables turn the Tolman--Oppenheimer--Volkoff equation with a scale-free equation of state such as $p=\gamma\rho$ into an autonomous system~\cite{Nilsson2001,Heinzle2003}, and in cosmology the approach is standard~\cite{Copeland1998}.

In the companion study~\cite{HBB} of the pure quadratic $f(G,B)$ model, the isotropy equation was found to be autonomous in the scale-invariant variables $U=r\mu'$, $V=r\nu'$. Two features of this model shape the present work. First, a regular centre forces the central density and pressure to vanish (Section~\ref{sec:centre}). For the affine equations of state used for compact stars a regular non-vacuum configuration is therefore impossible, and the non-vacuum configurations studied here have singular centres. Second, the isotropy equation is a \emph{single} first-order relation between the two functions $U$ and $V$, so a planar dynamical system requires a further relation. In~\cite{HBB} this was supplied by the minimum Euclidean norm closure, and the stability of the equilibrium curves was stated to be independent of that choice. We show that this holds on the two vacuum lines of the model, but that on the curve carrying the matter-bearing solutions the transverse eigenvalue can be given any value by a suitable closure (Section~\ref{sec:closure}). The physically meaningful closure is an equation of state, and it gives definite answers.

For a linear equation of state $p=\gamma\rho$ the closed system has three equilibria: a focus $P_1$, a saddle $P_2$, and a vacuum point $P_{\rm v}$ with nonzero bulk scalar, which exchanges stability with $P_2$ in a transcritical bifurcation at $\gamma=\tfrac23$. At the radiation value $\gamma=\tfrac13$ the system is integrable, and $P_1$ is a nonlinear centre surrounded by log-periodic solutions. The central result of this paper is the geometric consequence for light. The line in the $(U,V)$ plane on which the optical potential is stationary, and on which circular null orbits therefore lie, coincides at $\gamma=\tfrac13$, and only there, with the line containing the matter-bearing equilibria of the closure. Every periodic radiation core with monotonic areal radius has exactly one stable and one unstable light ring per period, and the radii of the two kinds of ring form two interlaced geometric progressions towards the centre with a common ratio (Proposition~\ref{prop:rings}). This structure is visible in the optical geometry, whose geodesics are the spatial tracks of light rays (Section~\ref{sec:optical}). There a self-similar solution appears as a cone without light rings, except at $\gamma=\tfrac13$, where the cone becomes a cylinder on which every parallel is a light ring, and a periodic core turns the cylinder into a periodic tube whose bulges and necks carry the stable and unstable rings. In the configurations of~\cite{KRS2000,KRS2001} the oscillations about the self-similar solution are not periodic; here the multiplicity is generated by an integrable radiation core, and the spacing is exactly geometric.

The field equations of the model were derived in~\cite{HBB,PLB}, and the autonomous form of the isotropy condition and the minimum norm closure were introduced in~\cite{HBB}. We do not repeat these derivations. Section~\ref{sec:model} collects the expressions we need; the remaining results go beyond that work.

The paper is organised as follows. Section~\ref{sec:model} summarises the model and the autonomous form of its isotropy condition, and Section~\ref{sec:centre} proves that a regular centre carries no matter. Section~\ref{sec:closure} describes the equilibria common to all closures of the isotropy condition and shows which stability statements survive a change of closure. Section~\ref{sec:eos} closes the system with a linear equation of state, Section~\ref{sec:radiation} treats the integrable radiation case, and Section~\ref{sec:lightrings} derives the nested light rings and describes them in the optical geometry. Section~\ref{sec:bag} discusses the MIT bag model and Section~\ref{sec:conclusion} concludes. The appendices contain the explicit closures (\ref{app:gauges}), the behaviour at infinity (\ref{app:infinity}) and the corresponding analysis in general relativity (\ref{app:GR}).

\section{The model and its autonomous isotropy condition}
\label{sec:model}

B\"ohmer and Jensko~\cite{BJ} split the Ricci scalar as $R=G+B$ into a bulk term $G$, quadratic in the Christoffel symbols, and a boundary term $B$, which is a total divergence, and considered gravitational actions built from a function $f(G,B)$. A function of $G$ alone gives field equations of at most second order, although this by itself does not establish the absence of ghosts or the physical viability of the theory. We restrict to $f=f(G)$, that is $f_B=0$, as a choice of model. Neither $G$ nor $B$ transforms as a scalar under general coordinate transformations, so a nonlinear $f(G)$ does not inherit general covariance and its field equations depend on the coordinate chart in which they are written~\cite{BJ}. The chart is therefore part of the model. Following~\cite{BJ,HBB,PLB}, we take the static spherically symmetric metric in isotropic coordinates of Cartesian form,
\begin{equation}
ds^{2} = -e^{\nu(r)}dt^{2} + e^{\mu(r)}\left(dx^{2}+dy^{2}+dz^{2}\right),
\label{eq:metric-cart}
\end{equation}
with $r=\sqrt{x^2+y^2+z^2}$, for which the bulk scalar is
\begin{equation}
G =\frac{1}{2}\,e^{-\mu}\mu'\left(\mu'+2\nu'\right),
\label{eq:G}
\end{equation}
a prime denoting $d/dr$. The spherical form of~(\ref{eq:metric-cart}) is used below only as geometric shorthand; the bulk scalar and the field equations are always those of the Cartesian chart.

Since $f=G$ reproduces general relativity, the simplest departure from Einstein gravity in this class is the pure quadratic model $f(G)=\epsilon G^{2}$, where $\epsilon$ has dimensions of $[\text{length}]^2$. Its field equations for a perfect fluid with energy density $\rho$ and pressure $p$, and the master equation expressing pressure isotropy, were derived in~\cite{HBB,PLB}. The $f(G)$ field equations do not imply a Bianchi identity in general~\cite{BJ}, so conservation is not automatic, but one verifies directly that these satisfy the conservation equation identically in the chart~(\ref{eq:metric-cart}); for isotropic pressure it reads $p'=-(\rho+p)\nu'/2$, as in general relativity. Once a pair $(\mu,\nu)$ satisfying the isotropy condition is known, $\rho$ and $p$ follow algebraically. A complete stellar model requires in addition an equation of state.

\subsection{Scale-invariant form}

The master isotropy equation of~\cite{HBB,PLB} is invariant under $r\mapsto\lambda r$, $\mu'\mapsto\lambda^{-1}\mu'$, $\nu'\mapsto\lambda^{-1}\nu'$ ($\lambda>0$). In the scale-invariant variables of~\cite{HBB},
\begin{equation}
U(X)=r\mu'(r),\qquad V(X)=r\nu'(r),\qquad X=\ln(r/r_0),
\label{eq:UV}
\end{equation}
with $r_0>0$ a fixed reference radius and a dot denoting $d/dX=r\,d/dr$, the explicit dependence on $r$ disappears and the isotropy condition becomes~\cite{HBB}
\begin{equation}
\boxed{\;\alpha(U,V)\,\dot U+\beta(U,V)\,\dot V=\Phi(U,V),\;}
\label{eq:constraint}
\end{equation}
where
\begin{align}
\alpha&=2\left(3U^2+6UV+2V^2\right), \qquad \beta=2U(3U+4V), \nonumber\\
\Phi&=U(U+2V)\,Q(U,V),\nonumber\\
Q&=3U^2+4UV+8U-V^2+8V. \label{eq:abPhi}
\end{align}
A radial rescaling becomes the translation $X\mapsto X+\ln\lambda$, which leaves~(\ref{eq:constraint}) unchanged. In the same variables, with $R=re^{\mu/2}$ the areal radius, the density and pressure take the scale-free form
\begin{equation}
\kappa\rho=\frac{\epsilon\,\mathcal{R}}{8R^4},\qquad \kappa p=\frac{\epsilon\,\mathcal{P}}{8R^4},
\label{eq:rhop-scalefree}
\end{equation}
with
\begin{align}
\mathcal{P}&=U(U+2V)\left(3U^2+6UV+8U+8V\right),\nonumber\\
\mathcal{R}&=U\big[U(U+2V)(5U-2V+8)\nonumber\\
&\qquad-(24U+32V)\dot U-16U\dot V\big].
\label{eq:PR}
\end{align}
The factor $R^{-4}$ reflects the dimension of $\epsilon$, and $d\ln R/dX=(U+2)/2$.

We write $\by=(U,V)$ and introduce the constraint normal and its rotation by a right angle,
\begin{equation}
\bn=(\alpha,\beta),\qquad \bnp=(-\beta,\alpha).
\label{eq:normals}
\end{equation}
For real $(U,V)$, $\bn$ vanishes only at the origin: $\beta=0$ requires $U=0$ or $V=-3U/4$, giving $\alpha=4V^2$ or $\alpha=-3U^2/4$ respectively. The general relativistic isotropy condition in isotropic coordinates also contains no dimensional constant, and in the same variables it becomes
\begin{equation}
\dot U+\dot V=2(U+V)+\tfrac12\left(U^2+2UV-V^2\right).
\label{eq:GRconstraint}
\end{equation}
Scale invariance of this kind underlies the homology-invariant dynamical systems formulations of relativistic stars~\cite{Nilsson2001,Heinzle2003}, where it is combined with a scale-free equation of state.

\section{No matter at a regular centre}
\label{sec:centre}

We consider equations of state of the affine form
\begin{equation}
p=\gamma(\rho-\rho_0),\qquad 0<\gamma\le1,
\label{eq:eos}
\end{equation}
which contains the linear equation of state ($\rho_0=0$) and the MIT bag model of quark matter~\cite{Chodos1974,Witten1984}, for which $\gamma=\tfrac13$ and $\rho_0=4B$ with $B$ the bag constant.

\begin{proposition}\label{prop:centre}
For every metric~(\ref{eq:metric-cart}) that is regular at the centre, the central density and pressure of the pure quadratic model vanish. With $\mu=\mu_0+\mu_2r^2+O(r^4)$ and $\nu=\nu_0+\nu_2r^2+O(r^4)$,
\begin{align}
\kappa\rho&=-40\epsilon e^{-2\mu_0}\mu_2^2(\mu_2+2\nu_2)\,r^2+O(r^4),\nonumber\\
\kappa p&=8\epsilon e^{-2\mu_0}\mu_2(\mu_2+\nu_2)(\mu_2+2\nu_2)\,r^2+O(r^4).
\label{eq:centre-expansion}
\end{align}
Consequently, for every affine equation of state~(\ref{eq:eos}) with $\gamma>0$, a configuration with a regular centre is vacuum.
\end{proposition}

\begin{proof}
With $U=2\mu_2r^2+O(r^4)$, $V=2\nu_2r^2+O(r^4)$, $\dot U=4\mu_2r^2+O(r^4)$ and $\dot V=4\nu_2r^2+O(r^4)$, direct expansion of~(\ref{eq:rhop-scalefree}) gives~(\ref{eq:centre-expansion}). For $\rho_0\neq0$ a regular centre would then require $p=-\gamma\rho_0\neq0$ at a point where $\rho=p=0$. For $\rho_0=0$ the conservation equation gives $p\,e^{(1+\gamma)\nu/(2\gamma)}=$ constant; since $\nu(0)$ is finite and $p(0)=0$, the constant vanishes, and $p=\rho=0$ throughout the connected chart.\qed
\end{proof}

In the $(U,V)$ plane a regular centre is the degenerate origin, where both potentials have vanishing logarithmic derivatives. The equation of state enters Proposition~\ref{prop:centre} only in its last step, and the isotropy condition alone restricts a regular centre further. Substituting the expansions into~(\ref{eq:constraint}), the lowest order, $r^6$, requires $\mu_2(\mu_2+\nu_2)(\mu_2+2\nu_2)=0$, so the $r^2$ term of $p$ in~(\ref{eq:centre-expansion}) vanishes and $p=O(r^4)$. On the branch $\nu_2=-\mu_2$ the density is $\kappa\rho=40\epsilon e^{-2\mu_0}\mu_2^3r^2+O(r^4)$, so that $p/\rho\to0$ at the centre. Proposition~\ref{prop:centre} therefore does not exclude regular centres for equations of state outside the class~(\ref{eq:eos}) with $p/\rho\to0$ as $\rho\to0$, such as polytropes, which we do not consider. The result is also specific to the pure quadratic model. Near a regular centre $G=2e^{-\mu_0}\mu_2(\mu_2+2\nu_2)r^2+O(r^4)$ tends to zero, so an added Einstein term $\alpha_0G$ dominates $\epsilon G^2$ there, and the conclusion does not carry over to $f=\alpha_0G+\epsilon G^2$ without a separate analysis. The non-vacuum configurations studied here therefore have singular centres, and the rest of the paper asks what structure such singular cores can have.

\section{Equilibria and the closure problem}
\label{sec:closure}

Equation~(\ref{eq:constraint}) is a single relation for two unknown functions. This mirrors the familiar fact that the isotropy condition alone does not determine both metric potentials, and it means that~(\ref{eq:constraint}) does not by itself define a vector field on the $(U,V)$ plane. At each point with $\bn\neq0$ the admissible velocities form the line $\bn\cdot\dot\by=\Phi$ in the velocity plane, and every such velocity can be written uniquely as
\begin{equation}
\dot\by=\Phi\,\frac{\bn}{|\bn|^2}+h\,\bnp,
\label{eq:allvel}
\end{equation}
with $h$ a free scalar. A choice of the function $h(U,V)$, which we call a closure or gauge, turns~(\ref{eq:constraint}) into a planar autonomous system. A constant pair $(U,V)$ corresponds to the self-similar metric
\begin{equation}
e^{\mu}=c_1\,r^{U},\qquad e^{\nu}=c_2\,r^{V},
\label{eq:powerlaw}
\end{equation}
which solves the isotropy condition exactly when $\Phi(U,V)=0$. We ask that every such exact solution be an orbit of the closed system, which forces $h$ to vanish wherever $\Phi$ does. Since $\Phi$ has simple zeros along its regular branches, this means $h=\Phi\,\Omega$ for a function $\Omega$ smooth away from the origin.

\begin{definition}
An \emph{admissible gauge} is a closure of~(\ref{eq:constraint}) of the form
\begin{equation}
\boxed{\;\dot\by=\Phi\left(\frac{\bn}{|\bn|^2}+\Omega(U,V)\,\bnp\right),\;}
\label{eq:gauge-general}
\end{equation}
with $\Omega$ smooth on an open set of the punctured plane.
\end{definition}

Equivalently, an admissible gauge is a smooth direction field $\bw(U,V)$ with $\bw\cdot\bn\neq0$, the closure being
\begin{equation}
\dot\by=\frac{\Phi}{\bw\cdot\bn}\,\bw,
\qquad
\Omega=\frac{\bw\cdot\bnp}{(\bw\cdot\bn)\,|\bn|^2}.
\label{eq:direction-gauge}
\end{equation}
The gauge is singular where $\bw\cdot\bn=0$, and the origin, where the constraint degenerates to $0=0$, is excluded in every gauge. A gauge is an auxiliary closure, not a choice of coordinates: different gauges select different families of metric potentials, and hence different inferred matter distributions. Correspondingly, stability in this section refers to the approach of neighbouring solutions to an equilibrium as the radius varies, not to temporal perturbations or to the pulsational stability of stars.

\subsection{Equilibria and their physical content}
\label{sec:physics}

In every admissible gauge the equilibrium set is $\{\Phi=0\}\setminus\{(0,0)\}$, consisting of
\begin{equation}
L_1:\ U=0,\qquad L_2:\ U+2V=0,\qquad C:\ Q=0.
\label{eq:curves}
\end{equation}
The conic $C$ has discriminant $4^2+4\cdot3=28>0$ and is a hyperbola through the origin. Solving $Q=0$ for $U$ gives two branches, each a graph over the whole $V$ axis,
\begin{equation}
C_{\pm}:\quad U=U_{\pm}(V)=\frac{-(2V+4)\pm\sqrt{7V^2-8V+16}}{3},
\label{eq:branches}
\end{equation}
since $7V^2-8V+16>0$ for all real $V$. The asymptotic directions of $C$ are $V/U=2\pm\sqrt7$. Apart from the origin, $C$ meets $L_1$ at $(0,8)$ on $C_+$ and $L_2$ at $(-\tfrac{16}{3},\tfrac{8}{3})$ on $C_-$. At these crossing points $\nabla\Phi=0$ and, by~(\ref{eq:Jacobian}) below, the Jacobian vanishes identically in every gauge. The origin corresponds to Minkowski space.

Each equilibrium point is an exact self-similar solution~(\ref{eq:powerlaw}). Substituting~(\ref{eq:powerlaw}) into~(\ref{eq:rhop-scalefree}), with $\dot U=\dot V=0$ and $R^4=c_1^2r^{2(U+2)}$, gives
\begin{align}
\kappa\rho&=\frac{\epsilon}{8c_1^2}\,U^2(U+2V)(5U-2V+8)\,r^{-2(U+2)}, \label{eq:rho-pl}\\
\kappa p&=\frac{\epsilon}{8c_1^2}\,U(U+2V)\nonumber\\
&\quad\times\left(3U^2+6UV+8U+8V\right)r^{-2(U+2)}. \label{eq:p-pl}
\end{align}
Since $G=U(U+2V)\,e^{-\mu}/(2r^2)$, the lines $L_1$ and $L_2$ are exactly the self-similar members of the set $G=0$, and on them $\rho=p=0$. More is true. For $\mu$ constant with $\nu$ arbitrary, and for $\mu=-2\nu+$ constant with $\nu$ arbitrary, the constraint~(\ref{eq:constraint}) holds identically and $\mathcal{R}=\mathcal{P}=0$, so every matter variable vanishes. These two vacuum families contain the vacuum solutions of~\cite{HBB},\footnote{In Sect.~6 of~\cite{HBB} the factorised vacuum condition and the two vacuum metrics are written with $\mu$ and $\nu$ interchanged relative to the line element used there. As printed, these metrics carry nonzero density and pressure; with the labels restored, they are members of the two families above.} where they were stated to be the only vacuum solutions of the model. They are not, as we now show.

On $C$ two identities hold,
\begin{align}
3U^2+6UV+8U+8V&=V(2U+V),\nonumber\\
U(5U-2V+8)&=(2U+V)(4U+8-V),
\label{eq:C-identities}
\end{align}
both following from $Q=0$. Hence the ratio of~(\ref{eq:p-pl}) to~(\ref{eq:rho-pl}) is
\begin{equation}
\gamma=\frac{V(2U+V)}{U(5U-2V+8)}=\frac{V}{4U+8-V}
\label{eq:gamma}
\end{equation}
on $Q=0$, where the second form cancels the common factor $2U+V$ and also follows from the conservation equation applied to the power laws. The cancellation fails where $2U+V=0$, that is at the origin and at the point
\begin{equation}
P_{\rm v}=\left(-\tfrac89,\ \tfrac{16}9\right)\in C_+ .
\label{eq:Pv}
\end{equation}
At $P_{\rm v}$ both left-hand sides in~(\ref{eq:C-identities}) vanish, so $\rho=p=0$, whereas
\begin{equation}
G=\frac{U(U+2V)}{2r^2e^{\mu}}=-\frac{32}{27R^2}\neq0 .
\end{equation}
$P_{\rm v}$ is therefore a vacuum solution with $G\neq0$, missed in the classification of~\cite{HBB}. In areal coordinates it is the curved metric
\begin{equation}
ds^2=-C\,R^{16/5}dt^2+\frac{81}{25}\,dR^2+R^2d\Omega^2,\qquad C>0,
\label{eq:Pv-metric}
\end{equation}
which describes the geometry of the Cartesian-chart solution; it is not a statement that the field equations hold in the areal chart. The same geometry is the power-law vacuum of $f(T)\propto T^2$ teleparallel gravity found by Golovnev and Guzm\'an~\cite{Golovnev2021}, whose exponent conditions are satisfied exactly by~(\ref{eq:Pv-metric}); the relation between the bulk and teleparallel formulations is discussed in~\cite{BJ}. What the present setting adds is its role in the matter dynamics: $P_{\rm v}$ is an equilibrium of every closure by a linear equation of state, and it exchanges stability with the matter-bearing equilibrium $P_2$ at $\gamma=\tfrac23$, exactly where the density of $P_2$ changes sign (Section~\ref{sec:eos}).

Besides $P_{\rm v}$ and the origin, $C$ contains three exceptional points. At its crossings with $L_1$ and $L_2$ the density and pressure vanish, and at $(-\tfrac83,-\tfrac83)\in C_-$, where $4U+8-V=0$, the density vanishes while $\mathcal{P}=4096/9$, so that no finite $\gamma$ applies. Every other point of $C$ is an exact perfect-fluid solution with linear equation of state $p=\gamma\rho$, singular at the centre, where $\rho\propto R^{-4}$ diverges.

\subsection{Which stability statements survive a change of closure}

\begin{proposition}\label{prop:lin}
Let $p$ be a point of $\Phi=0$ with $\bn(p)\neq0$, and consider an admissible gauge with direction field $\bw$, $\bw\cdot\bn\neq0$ at $p$. The Jacobian of~(\ref{eq:direction-gauge}) at $p$ is
\begin{equation}
J_p=\frac{1}{\bw\cdot\bn}\;\bw\,(\nabla\Phi)^{\mathsf T}\Big|_p,
\label{eq:Jacobian}
\end{equation}
of rank at most one. If $p$ is a regular point of $\Phi=0$, its eigenvalues are $\lambda_\parallel=0$, with eigenvector tangent to the equilibrium curve, and
\begin{equation}
\lambda_\perp=\frac{\bw\cdot\nabla\Phi}{\bw\cdot\bn}\Big|_p=\lmin+\Omega\,W,\qquad
\lmin=\frac{D}{|\bn|^2},
\label{eq:lambda-omega}
\end{equation}
where
\begin{align}
D&=\bn\cdot\nabla\Phi=\alpha\Phi_U+\beta\Phi_V,\nonumber\\
W&=\bnp\cdot\nabla\Phi=\alpha\Phi_V-\beta\Phi_U.
\label{eq:DW}
\end{align}
\end{proposition}

\begin{proof}
Writing $\dot\by=\Phi\,\bw/(\bw\cdot\bn)$, every term of the derivative that does not differentiate $\Phi$ carries a factor $\Phi$ and vanishes at $p$, which gives~(\ref{eq:Jacobian}). A rank-one matrix $\mathbf{a}\mathbf{b}^{\mathsf T}$ has eigenvalues $0$ and $\mathbf{b}\cdot\mathbf{a}$. For the $\Omega$ form take $\bw=\bn/|\bn|^2+\Omega\bnp$, for which $\bw\cdot\bn=1$.\qed
\end{proof}

Since $X=\ln r$, $\lambda_\perp<0$ means that neighbouring solutions approach the equilibrium curve as $r$ increases, and $\lambda_\perp>0$ that they leave it. On $L_1$ and $L_2$ the vectors $\bn$ and $\nabla\Phi$ are parallel, so $W=0$ and
\begin{equation}
L_1:\ \lambda_\perp=\frac{8-V}{2},\qquad L_2:\ \lambda_\perp=\frac{3U+16}{4}
\label{eq:L12}
\end{equation}
in every admissible gauge. On $C$ one finds, with $\nabla\Phi=U(U+2V)\nabla Q$,
\begin{align}
W\big|_C&=-4U(U+2V)\big(3U^3+9U^2V+10UV^2\nonumber\\
&\qquad-8UV+2V^3-8V^2\big).
\label{eq:WC}
\end{align}
Eliminating $U$ between $Q=0$ and the cubic factor gives the resultant $-96V^3(63V^3-520V^2+1088V-1024)$, whose cubic has a single real root. Apart from the origin and the crossing points, $W$ therefore vanishes on $C$ only at
\begin{equation}
p_*=(U_*,V_*)\approx(-0.437952,\ 5.737858)\in C_+,
\label{eq:pstar}
\end{equation}
where a Gr\"obner basis computation gives $\lambda_\perp=-\tfrac43$ exactly. At $p_*$,~(\ref{eq:rho-pl}) and~(\ref{eq:p-pl}) give $\rho$ and $p$ with the sign of $-\epsilon$ and $p/\rho\approx11.24$, so the point is unphysical for either sign of $\epsilon$.

\begin{theorem}[Gauge dependence of transverse stability]\label{thm:main}
Let $p$ be a regular point of $\Phi=0$ other than the origin.
\begin{enumerate}[label=(\roman*)]
\item If $W(p)=0$, then $\lambda_\perp(p)=D(p)/|\bn(p)|^2$ in every admissible gauge.
\item If $W(p)\neq0$, then for every real number $c$ there is an admissible gauge, namely the constant $\Omega=(c-\lmin(p))/W(p)$, with $\lambda_\perp(p)=c$.
\item Let $K$ be a compact arc of $\Phi=0$ containing no zero of $W$. Then there exist constants $\Omega_-$ and $\Omega_+$ such that $K$ is transversely attracting at every point in the gauge $\Omega=\Omega_-$ and transversely repelling at every point in the gauge $\Omega=\Omega_+$.
\end{enumerate}
\end{theorem}

\begin{proof}
Parts (i) and (ii) are immediate from~(\ref{eq:lambda-omega}). For (iii), $W$ has a fixed sign on $K$ and $|W|$ is bounded below there by some $w_0>0$, while $|\lmin|\le M$ on $K$. If $W>0$ on $K$, then $\Omega_-<-M/w_0$ and $\Omega_+>M/w_0$ suffice; if $W<0$, interchange them.\qed
\end{proof}

\begin{figure*}[t]
\centering
\includegraphics[width=0.86\textwidth]{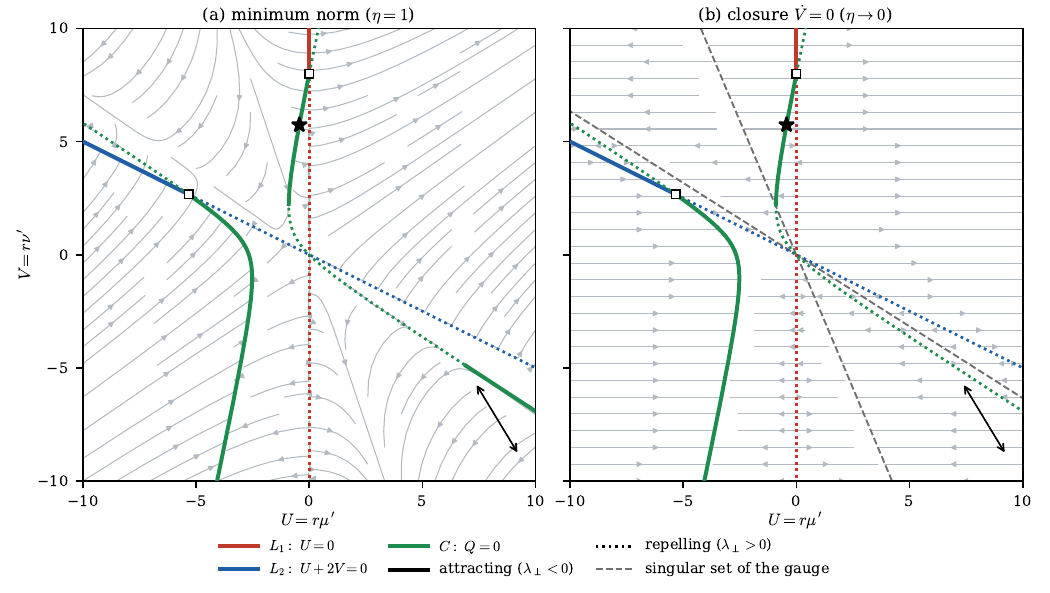}
\caption{The two closures of the isotropy condition used in~\cite{HBB}: (a) the minimum norm closure and (b) the closure $\dot V=0$, which are the members $\eta=1$ and $\eta\to0$ of the weighted family of \ref{app:gauges}. The equilibrium set is the same: the vacuum lines $L_1$ (red) and $L_2$ (blue) and the hyperbola $C$ (green), drawn solid where transversely attracting and dotted where repelling. The classification of $L_1$ and $L_2$ agrees; the arc $V<-4.87$ of $C_+$ (double arrow) is attracting in (a) and repelling in (b). Open squares mark the crossing points and the star the point $p_*$; dashed grey lines in (b) are the lines $\alpha=0$ on which that closure is singular}
\label{fig:pair}
\end{figure*}

The transverse stability of the vacuum lines is thus a property of the field equations, given by~(\ref{eq:L12}), while that of every point of the matter-bearing hyperbola other than $p_*$ is a property of the closure alone. The geometric reason is simple: the constraint fixes only the component of the velocity along $\bn$, whereas the transverse growth rate of $\Phi$ depends on the component along $\nabla\Phi$, and the closure supplies the missing component. The two closures used in~\cite{HBB} illustrate the point (Fig.~\ref{fig:pair}). The minimum norm closure, $\bw=\bn$, makes the arc $V<-4.8662$ of $C_+$ attracting. The closure $\dot V=0$, $\dot U=\Phi/\alpha$, which was used there as a check, has transverse eigenvalue $\Phi_U/\alpha$; this has no sign change on $C_+$ for $V<0$, so the same arc is repelling, with $\Phi_U/\alpha\approx32.85$ at $V=-6$ against $\lmin\approx-8.06$. Conversely, agreement between particular closures is no evidence of invariance. The gradient normal closure $\bw=\nabla\Phi$ reproduces the minimum norm classification at every regular point, but only because both transverse eigenvalues, $D/|\bn|^2$ and $|\nabla\Phi|^2/D$, have the sign of $D$; it is the criterion $W=0$ that separates properties of the field equations from properties of the closure. Both closures of~\cite{HBB} are members of the one-parameter family $\bw=(\alpha,\eta\beta)$, at $\eta=1$ and in the limit $\eta\to0$, which is studied together with other explicit closures in \ref{app:gauges}; the equilibria at infinity are equally closure dependent (\ref{app:infinity}). The same mechanism operates in general relativity, where the only self-similar solutions with closure-independent transverse stability are flat (\ref{app:GR}). The isotropy condition alone therefore cannot assign a stability to the matter-bearing equilibria, and a physical closure is needed.

\section{Closure by a linear equation of state}
\label{sec:eos}

The missing relation has a physical counterpart: an equation of state. An equation of state is a closure of the form~(\ref{eq:allvel}), but not an admissible gauge: it does not leave every point of $\Phi=0$ an equilibrium. Instead it singles out isolated equilibria, whose stability is then fixed by the field equations and the matter model alone.

\subsection{The equation of state as a second relation}

By~(\ref{eq:rhop-scalefree}), the equation of state~(\ref{eq:eos}) becomes $\mathcal{P}-\gamma\mathcal{R}=-\gamma\Lambda$, where
\begin{equation}
\Lambda=\frac{8\kappa\rho_0R^4}{\epsilon},\qquad \dot\Lambda=2(U+2)\Lambda,
\label{eq:Lambda}
\end{equation}
the second relation following from $R^4=r^4e^{2\mu}$. Written out, the equation of state is a second relation linear in the derivatives,
\begin{multline}
8\gamma U\left[(3U+4V)\dot U+2U\dot V\right]=U(U+2V)\\
\times\left[\gamma U(5U-2V+8)-\left(3U^2+6UV+8U+8V\right)\right]\\
-\gamma\Lambda .
\label{eq:eos-constraint}
\end{multline}
Together with the isotropy constraint~(\ref{eq:constraint}) it forms a linear system for $(\dot U,\dot V)$ with determinant $-48\gamma U^2(U+2V)^2$. The system is therefore uniquely solvable away from the vacuum lines, on which $\rho$ and $p$ vanish identically and the equation of state carries no information. Solving it gives an autonomous system in $(U,V,\Lambda)$. For $\rho_0=0$ this reduces to the plane $\Lambda=0$. For $\rho_0\neq0$ the plane $\Lambda=0$ remains invariant and describes the limit $R\to0$, in which the density grows without bound and the bag constant becomes negligible.

As a consistency check, the closed system for $\rho_0=0$ preserves the Tolman integral: $\kappa p\,e^{(1+\gamma)\nu/(2\gamma)}$ is constant along every orbit, as the conservation equation requires. One consequence is that the zero-pressure curve $3U^2+6UV+8U+8V=0$ is invariant when $\rho_0=0$, so a linear equation of state never produces a stellar surface.

\subsection{Equilibria and the transcritical bifurcation}

For $\rho_0=0$ the equilibria of the closed system are the points of the hyperbola $C$ at which the right side of~(\ref{eq:eos-constraint}) vanishes. By~(\ref{eq:C-identities}), on $Q=0$ this condition factorises as
\begin{equation}
(2U+V)\left[(1+\gamma)V-4\gamma(U+2)\right]=0.
\label{eq:eos-eq}
\end{equation}
The second factor gives the straight line
\begin{equation}
V=k(U+2),\qquad k=\frac{4\gamma}{1+\gamma},
\label{eq:eos-line}
\end{equation}
through the point $(-2,0)$. Substituting into $Q=0$ and writing $a=U+2$ gives
\begin{align}
&(3+4k-k^2)a^2-4a-4=0,\qquad s=\sqrt{4+4k-k^2},\nonumber\\
&a_1=-\frac{2}{s+1},\qquad a_2=\frac{2}{s-1},
\label{eq:eos-roots}
\end{align}
with $s>2$ for $0<k<4$, that is for every $\gamma>0$. This gives the equilibria $P_1=(a_1-2,\,ka_1)$ on $C_-$ and $P_2=(a_2-2,\,ka_2)$ on $C_+$. The first factor of~(\ref{eq:eos-eq}) gives the vacuum point $P_{\rm v}$ of~(\ref{eq:Pv}) for every $\gamma$. The closure is regular there, since its determinant does not vanish, so $P_{\rm v}$ is an equilibrium of the closed system although it carries no matter. There are therefore three equilibria for every $\gamma>0$, except at $\gamma=\tfrac23$, where $P_2=P_{\rm v}$ (Table~\ref{tab:eos}). If only strictly positive densities are admitted, $P_{\rm v}$ is a boundary point of the physical region.

At $P_1$ and $P_2$,~(\ref{eq:rho-pl}) gives $\kappa\rho\propto\epsilon(U+2V)(5U-2V+8)$. This factor is positive at $P_1$ for all $\gamma\in(0,1]$ and at $P_2$ for $\gamma<\tfrac23$, so we take $\epsilon>0$. At $P_1$ one has $U+2<0$, so $R\propto r^{(U+2)/2}$ decreases as $r$ increases, and the direction $X\to+\infty$ is the centre $R\to0$; at $P_2$ and $P_{\rm v}$, $U+2>0$.

\begin{figure}[t]
\centering
\includegraphics[width=\columnwidth]{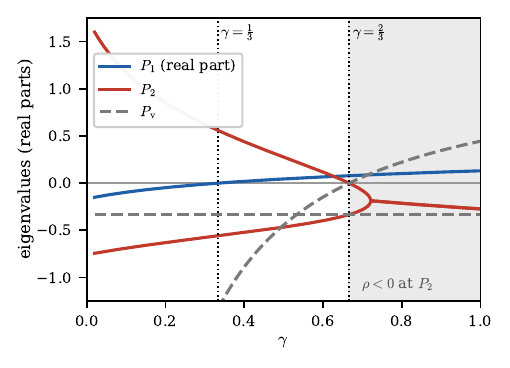}
\caption{Real parts of the eigenvalues at $P_1$, $P_2$ and $P_{\rm v}$ as functions of $\gamma$. The focus $P_1$ changes stability at the radiation value $\gamma=\tfrac13$. $P_2$ and $P_{\rm v}$ exchange stability in a transcritical bifurcation at $\gamma=\tfrac23$, beyond which the density at $P_2$ is negative (shaded)}
\label{fig:eos-bif}
\end{figure}

The stability of all three equilibria follows in closed form. Reducing the Jacobian modulo the quadratic in~(\ref{eq:eos-roots}) gives, at $P_1$ and $P_2$,
\begin{align}
\operatorname{tr}J&=\frac{(1-3\gamma)(U+2)}{2(1+\gamma)},\nonumber\\
\det J&=\frac{2(k+4)-(2k+1)(4-k)a}{3k}.
\label{eq:trdet}
\end{align}
At $P_1$, where $a_1<0$, one has $\det J>2(k+4)/(3k)\ge2$ and $|\operatorname{tr}J|<\tfrac13$ for $0<\gamma\le1$, so the eigenvalues form a complex pair. $P_1$ is therefore a focus, attracting as $X$ increases for $\gamma<\tfrac13$ and repelling for $\gamma>\tfrac13$, and its trace vanishes exactly at radiation. At $P_2$,
\begin{equation}
\det J=\frac{2s\,(k+4-2s)}{3k(s-1)},
\end{equation}
which has the sign of $5k^2-8k$, that is of $3\gamma-2$: $P_2$ is a saddle for $\gamma<\tfrac23$ and stable for $\gamma>\tfrac23$, where its density is negative. At $P_{\rm v}$ the eigenvalues are
\begin{equation}
\lambda_1=-\frac13,\qquad \lambda_2=\frac{4(3\gamma-2)}{9\gamma},
\label{eq:Pv-eig}
\end{equation}
so $P_{\rm v}$ is a sink for $\gamma<\tfrac23$ and a saddle for $\gamma>\tfrac23$. $P_2$ and $P_{\rm v}$ thus exchange stability as they pass through each other at $\gamma=\tfrac23$. With $q=(1,-11)$ the centre eigenvector there and $x$ the coordinate along $q$ in the eigenbasis at $P_{\rm v}$, the flow on the centre manifold is
\begin{equation}
\dot x=2\left(\gamma-\tfrac23\right)x+\tfrac{63}{4}\,x^2+\cdots.
\label{eq:transcritical}
\end{equation}
Both coefficients are nonzero, so the exchange is a non-degenerate transcritical bifurcation (Fig.~\ref{fig:eos-bif}).

In terms of the areal radius, the singular self-similar solution $P_1$ is the attractor of the core, $R\to0$, when $\gamma<\tfrac13$. When $\gamma>\tfrac13$ it is the attractor at large $R$, which is the behaviour familiar from the general relativistic isothermal sphere~\cite{Nilsson2001}. For $\gamma<\tfrac23$ the vacuum point $P_{\rm v}$ is a sink, and since $U+2=\tfrac{10}9>0$ there, it attracts nearby solutions as the areal radius increases. The radiation value $\gamma=\tfrac13$ separates the two regimes of $P_1$, and there the system is integrable.

\begin{table*}[t]
\centering
\caption{Equilibria of the closure by the linear equation of state $p=\gamma\rho$, with eigenvalues with respect to $X=\ln r$. The vacuum point $P_{\rm v}=(-\tfrac89,\tfrac{16}9)$ is an equilibrium for every $\gamma$, with eigenvalues~(\ref{eq:Pv-eig}); at $\gamma=\tfrac23$ it coincides with $P_2$. The density is positive for $\epsilon>0$ at $P_1$ throughout and at $P_2$ for $\gamma<\tfrac23$}
\label{tab:eos}
\small
\begin{tabular}{cccccc}
\toprule
$\gamma$ & $P_1$ & eigenvalues at $P_1$ & $P_2$ & eigenvalues at $P_2$ & eigenvalues at $P_{\rm v}$\\
\midrule
$0.1$ & $(-2.6048,\,-0.2199)$ & $-0.0962\pm3.3871\,i$ & $(-0.4698,\,0.5564)$ & $-0.6890,\ 1.1759$ & $-0.3333,\ -7.5556$\\
$0.25$ & $(-2.5616,\,-0.4493)$ & $-0.0281\pm2.4385\,i$ & $(-0.7190,\,1.0248)$ & $-0.6030,\ 0.7311$ & $-0.3333,\ -2.2222$\\
$1/3$ & $(-2.5486,\,-0.5486)$ & $\pm2.2314\,i$ & $(-0.7847,\,1.2153)$ & $\pm0.5589$ & $-0.3333,\ -1.3333$\\
$0.5$ & $(-2.5335,\,-0.7114)$ & $0.0445\pm1.9922\,i$ & $(-0.8563,\,1.5249)$ & $-0.4669,\ 0.2763$ & $-0.3333,\ -0.4444$\\
$2/3$ & $(-2.5263,\,-0.8421)$ & $0.0789\pm1.8527\,i$ & $=P_{\rm v}$ & $-0.3333,\ 0$ & $-0.3333,\ 0$\\
$1$ & $(-2.5224,\,-1.0448)$ & $0.1306\pm1.6893\,i$ & $(-0.9062,\,2.1877)$ & $-0.2735\pm0.3196\,i$ & $-0.3333,\ 0.4444$\\
\bottomrule
\end{tabular}
\end{table*}

\begin{figure*}[t]
\centering
\includegraphics[width=\textwidth]{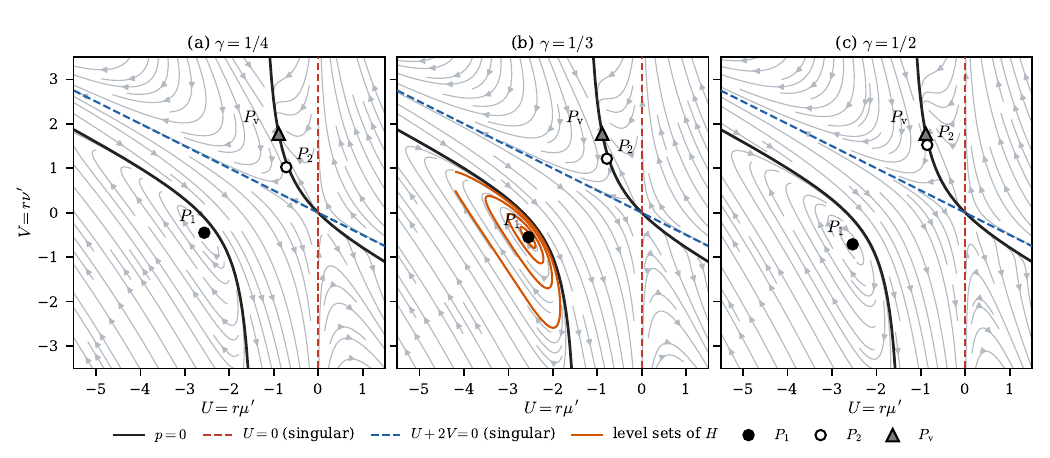}
\caption{Closure by the linear equation of state $p=\gamma\rho$ for (a) $\gamma=\tfrac14$, (b) $\gamma=\tfrac13$ and (c) $\gamma=\tfrac12$. The focus $P_1$ is stable, a centre and unstable respectively, $P_2$ is a saddle, and the vacuum point $P_{\rm v}$ (triangle) is a sink. In (b) level sets of the first integral~(\ref{eq:H}) are shown. The vacuum lines (dashed) are singular for this closure, and the zero-pressure curve (solid black) is invariant}
\label{fig:eos}
\end{figure*}

\section{Radiation cores}
\label{sec:radiation}

\begin{theorem}\label{thm:radiation}
For $\gamma=\tfrac13$ the closed system has the first integral
\begin{equation}
H(U,V)=\frac{3U^2+6UV+8U+8V}{U^3(U+2V)^3(2U+V)^4}
\label{eq:H}
\end{equation}
on the domain $U(U+2V)(2U+V)\neq0$. The equilibria in this domain are $P_{1,2}=\big((-5\mp\sqrt7)/3,\,(1\mp\sqrt7)/3\big)$, with eigenvalues $\pm\tfrac{i}{3}\sqrt{21+9\sqrt7}$ at $P_1$ and $\pm\tfrac13\sqrt{9\sqrt7-21}$ at $P_2$. The third equilibrium, $P_{\rm v}$, lies on $2U+V=0$, outside the domain of $H$, and is a sink with eigenvalues $-\tfrac13$ and $-\tfrac43$. $H$ has a non-degenerate local maximum at $P_1$, which is therefore a nonlinear centre surrounded by a period annulus of closed orbits, the level sets of $H$.
\end{theorem}

The integral has a transparent origin. For $\gamma=\tfrac13$ the Tolman integral reads $\kappa p\,e^{2\nu}=$ constant. In addition, the quantity
\begin{equation}
N=\frac{e^{(\nu-\mu)/2}}{r}\,U(U+2V)(2U+V)
\end{equation}
is conserved, as one verifies directly from the closed system. It is the Noether charge of the global rescaling $g_{ab}\to e^{2\sigma}g_{ab}$, which leaves $\int G^2\sqrt{-g}\,d^4x$ invariant and is respected by a traceless fluid. Both conserved quantities depend on the potentials themselves, but their scale-invariant combination does not: $\kappa p\,e^{2\nu}/N^4=\epsilon H/8$. For $\gamma\neq\tfrac13$ the trace of the matter breaks the symmetry and $N$ is not conserved. Correspondingly, the line $2U+V=0$, which is invariant at $\gamma=\tfrac13$, is crossed by the flow at a rate proportional to $3\gamma-1$.

The centre establishes a local family of log-periodic radiation configurations; it does not imply that every radiation solution is periodic. Likewise, the change of stability of $P_1$ at $\gamma=\tfrac13$ is an integrable centre--focus transition rather than a generic Hopf bifurcation, and no limit cycle is implied.

Each closed orbit is a solution periodic in $\ln r$, that is, a log-periodic, discretely self-similar fluid. Since $U+2<0$ near $P_1$, increasing $X$ is the direction towards the centre. In the small-amplitude limit the period in $X$ is $6\pi/\sqrt{21+9\sqrt7}\approx2.816$, and over one period, traversed towards increasing $X$, the areal radius changes by
\begin{equation}
\Delta\ln R=\frac{\pi(1-\sqrt7)}{\sqrt{21+9\sqrt7}}\approx-0.772 .
\label{eq:DeltalnR}
\end{equation}
The radius thus contracts by a factor of about $0.462$ per oscillation towards the centre, or grows by a factor of about $2.165$ per oscillation traversed outwards, while the density scales as $R^{-4}$ modulated by a periodic factor. These values are small-amplitude limits: the period grows with amplitude, and we verified numerically that orbits close down to $H\approx0.18\,H(P_1)$. The closed orbits that lie in the half-plane $U<-2$, along which the areal radius is monotone, are those with $H>H_c\approx0.576\,H(P_1)$, where $H_c$ is the maximum of $H$ on the part of the line $U=-2$ below $L_2$, attained at $V=(1-\sqrt{31})/3$. They include all the orbits used below.

\section{Nested light rings}
\label{sec:lightrings}

\subsection{The light-ring line}

Photons follow null geodesics of the metric~(\ref{eq:metric-cart}), to which the fluid couples minimally. For a static spherically symmetric metric written in areal form, $ds^2=-e^{\nu}dt^2+g_{RR}\,dR^2+R^2d\Omega^2$, the conserved energy $E$ and angular momentum $L$ of a null geodesic give
\begin{equation}
e^{\nu}g_{RR}\left(\frac{dR}{d\lambda}\right)^2=E^2-L^2\,\Vopt,\qquad \Vopt=\frac{e^{\nu}}{R^2},
\label{eq:null}
\end{equation}
with $\lambda$ an affine parameter. Circular null orbits lie at stationary points of the optical potential $\Vopt$: a maximum gives a radially unstable light ring and a minimum a radially stable one~\cite{Cunha2017}. Maxima of $\Vopt$ are the necks of the optical geometry~\cite{Abramowicz1988}, the spatial metric $e^{-\nu}(g_{RR}dR^2+R^2d\Omega^2)$ whose geodesics are the spatial tracks of null geodesics, and minima are its bulges~\cite{Abramowicz1997,KRS2001}; Section~\ref{sec:optical} develops this picture for the radiation cores. In the variables~(\ref{eq:UV}),
\begin{equation}
\frac{d\ln\Vopt}{dX}=V-U-2,
\label{eq:dVopt}
\end{equation}
so wherever $U+2\neq0$, and the areal radius is therefore monotone in $X$, the light rings are the points at which a solution crosses the straight line
\begin{equation}
V=U+2 .
\label{eq:LR-line}
\end{equation}

Compare~(\ref{eq:LR-line}) with the line~(\ref{eq:eos-line}) on which the closure by $p=\gamma\rho$ has its matter-bearing equilibria. The two coincide if and only if $k=1$, that is $\gamma=\tfrac13$. On the line~(\ref{eq:eos-line}), $d\ln\Vopt/d\ln R=2(3\gamma-1)/(1+\gamma)$, so the optical potential of the self-similar solution $P_1$ is a power of $R$ that is decreasing for $\gamma<\tfrac13$, increasing for $\gamma>\tfrac13$, and constant only at radiation. The exact self-similar radiation solution therefore admits a circular null orbit at every radius. The origin of the special value is the Tolman scaling: a self-similar fluid has $e^{\nu}\propto\rho^{-2\gamma/(1+\gamma)}$, and since $\rho\propto R^{-4}$ in the quadratic model, the optical potential is constant when $\gamma=\tfrac13$. In general relativity, where self-similar fluids have $\rho\propto R^{-2}$, the same argument selects the stiff value $\gamma=1$.

\begin{figure*}[t]
\centering
\includegraphics[width=\textwidth]{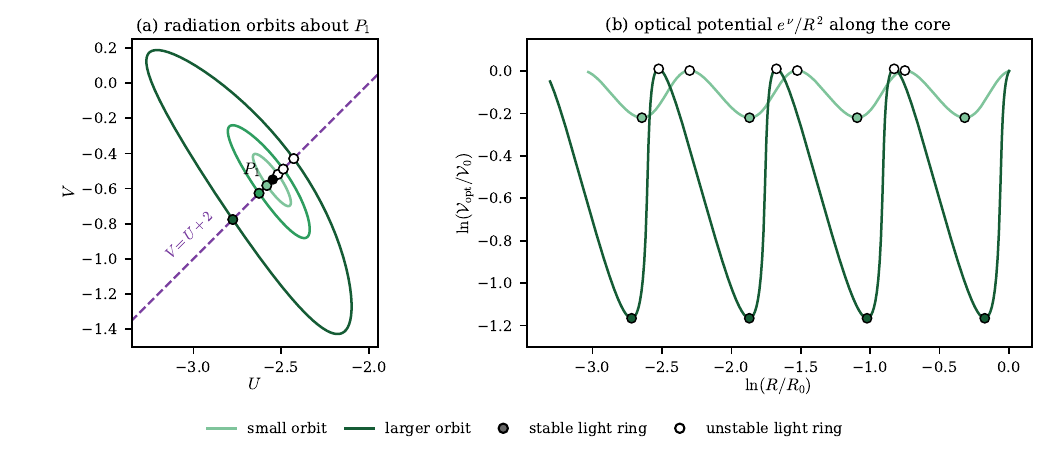}
\caption{Light rings of radiation cores ($\gamma=\tfrac13$). (a) Closed orbits of the radiation system about $P_1$ for initial displacements $0.05$, $0.1$ and $0.2$ along $U$, with the light-ring line $V=U+2$ (dashed). Each orbit crosses the line once on each side of $P_1$ per period, at a radially stable (filled) light ring on the left and a radially unstable (open) one on the right. (b) The optical potential $e^{\nu}/R^2$ along two of these cores against the areal radius. It is periodic in $\ln R$, so the stable and unstable rings form two interlaced geometric progressions with a common ratio, here $2.171$ (light curve) and $2.335$ (dark curve)}
\label{fig:rings}
\end{figure*}

\subsection{Light rings of periodic radiation cores}

\begin{proposition}\label{prop:rings}
Let $\gamma=\tfrac13$ and consider a closed orbit of the period annulus of $P_1$ that lies in the half-plane $U<-2$. Along the corresponding solution,
\begin{equation}
\Vopt\propto\left(pR^4\right)^{-1/2}\propto\mathcal{P}(U,V)^{-1/2},
\label{eq:Vopt-P}
\end{equation}
so the optical potential is a periodic function of $X$. In each period the orbit crosses the line~(\ref{eq:LR-line}) exactly twice, transversally and on opposite sides of $P_1$: for $U<U_1=(-5-\sqrt7)/3$ at a minimum of $\Vopt$, a radially stable light ring, and for $U>U_1$ at a maximum, a radially unstable one. The solution therefore contains infinitely many light rings accumulating at the centre, alternately stable and unstable. The radii of the stable rings and those of the unstable rings form two interlaced geometric progressions with the common ratio $e^{|\Delta\ln R|}$, where $\Delta\ln R=\oint(U+2)\,dX/2$ is the change of $\ln R$ over one period. In the small-amplitude limit this ratio is
\begin{equation}
\exp\!\left(\frac{\pi(\sqrt7-1)}{\sqrt{21+9\sqrt7}}\right)\approx2.165 .
\label{eq:ratio}
\end{equation}
\end{proposition}

\begin{proof}
The Tolman integral at $\gamma=\tfrac13$ gives $e^{\nu}\propto p^{-1/2}$, hence $\Vopt=e^{\nu}/R^2\propto(pR^4)^{-1/2}$, and $pR^4=\epsilon\mathcal{P}/(8\kappa)$ by~(\ref{eq:rhop-scalefree}). Since $\mathcal{P}$ depends on $(U,V)$ only, $\Vopt$ returns to the same value after each period, while $\ln R$ changes by $\Delta\ln R$. Equivalently, the closed system satisfies the identity $d\ln\mathcal{P}/dX=-2(V-U-2)$, consistent with~(\ref{eq:dVopt}). Write $D=V-U-2$, so that $D=d\ln\Vopt/dX$. On the line $D=0$ the closed system gives
\begin{equation}
\dot D=\frac{(3U^2+10U+6)(27U^2+36U+8)}{3U(3U+4)} .
\label{eq:Ddot}
\end{equation}
For $U<-2$ the denominator and the second factor of the numerator are positive, while the first factor, with roots $U_1$ and $(-5+\sqrt7)/3$, is positive for $U<U_1$ and negative for $U_1<U<-2$. Every crossing of the orbit with~(\ref{eq:LR-line}) is therefore transverse: $D$ increases through zero to the left of $P_1$, giving a minimum of $\Vopt$ in $X$, and decreases to its right, giving a maximum. Since $R$ is monotone in $X$, these are a minimum and a maximum in $R$ as well. The orbit is a simple closed curve surrounding $P_1$, so its winding number about $P_1$ is $\pm1$, and this equals its signed number of crossings with each of the two half-lines into which $P_1$ divides~(\ref{eq:LR-line}). All crossings with a given half-line have the same sign, so the orbit crosses each half-line exactly once per period. The stable and unstable rings therefore alternate, and each type recurs once per period at radii reduced by the factor $e^{\Delta\ln R}$. The limit~(\ref{eq:ratio}) follows from~(\ref{eq:DeltalnR}).\qed
\end{proof}

The identity $d\ln\mathcal{P}/dX=-2(V-U-2)$ and~(\ref{eq:Ddot}) were verified symbolically from the closed system. By Section~\ref{sec:radiation}, Proposition~\ref{prop:rings} applies to every closed orbit with $H>H_c$. Numerically, for orbits through $P_1+(s,0)$ the ratio is $2.171$, $2.193$, $2.240$ and $2.335$ for $s=0.05$, $0.1$, $0.15$ and $0.2$, that is for $H/H(P_1)\approx0.988$, $0.946$, $0.862$ and $0.722$, constant along each orbit to better than $10^{-10}$. The factors between adjacent rings alternate. For $s=0.1$, each unstable ring lies at $1.349$ times the radius of the stable ring immediately inside it, and the next stable ring outwards lies a further factor $1.626$ out, the product of the two being $2.193$. In the small-amplitude limit both factors tend to $\sqrt{2.165}\approx1.471$, and the two progressions become evenly interlaced. Figure~\ref{fig:rings} shows the crossings and the optical potential along two cores.

For $\gamma\neq\tfrac13$ the equilibrium $P_1$ lies off the line~(\ref{eq:LR-line}), at a distance proportional to $|3\gamma-1|$. Solutions that approach the focus $P_1$ therefore cross the line only finitely often, and the infinite nested sequence of light rings is special to the radiation value.

\begin{figure*}[t]
\centering
\includegraphics[width=\textwidth]{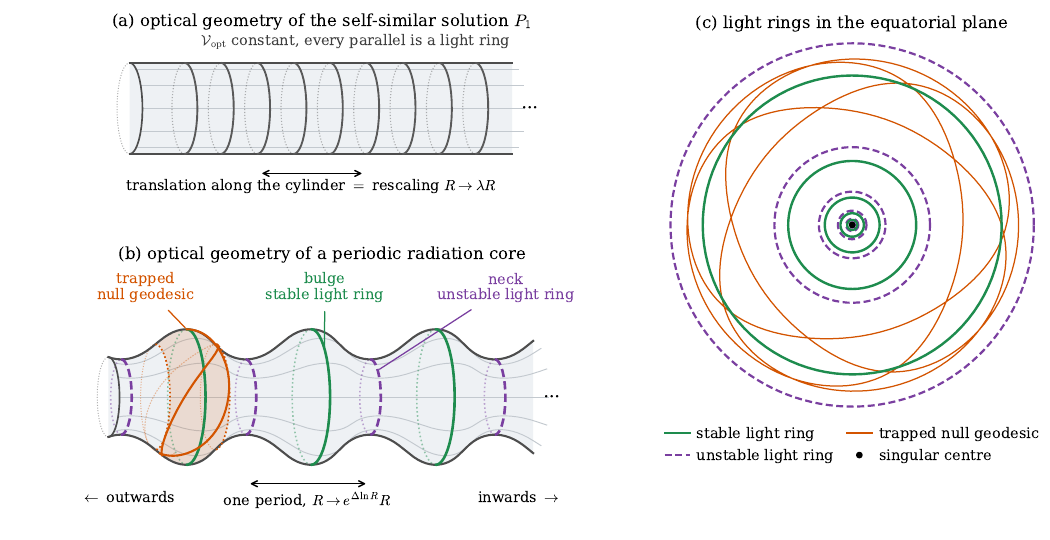}
\caption{Optical geometry of radiation cores ($\gamma=\tfrac13$) in the equatorial plane, drawn as surfaces of revolution in Euclidean space on which null geodesics are geodesics. Each surface is fixed up to an overall scale. (a) For the self-similar solution $P_1$ the surface is a cylinder, every parallel is a circular null orbit, and a translation along the cylinder is a rescaling of the solution. (b) For the core through $P_1+(0.2,0)$, the larger orbit of Fig.~\ref{fig:rings}, the surface is a periodic tube whose necks carry the unstable light rings (dashed) and whose bulges carry the stable ones (solid). A null geodesic with impact parameter $b=0.78\,\varrho_{\rm bulge}$, where $\varrho_{\rm neck}=0.56\,\varrho_{\rm bulge}$, is trapped in the shaded band about a bulge, between the parallels $\varrho=b$ (dotted); the segment drawn completes two radial oscillations. Translation by one period is the rescaling $R\to e^{\Delta\ln R}R$, with $e^{|\Delta\ln R|}=2.335$, and the centre lies at infinite optical distance to the right. (c) The same light rings, and the trapped geodesic followed over five radial oscillations, in the equatorial plane in areal radius, normalised to the outermost unstable ring shown. Each unstable ring lies at $1.216$ times the radius of the stable ring inside it, and the rings of each type accumulate at the singular centre with ratio $2.335$}
\label{fig:optical}
\end{figure*}

\subsection{The optical geometry of the cores}
\label{sec:optical}

In the equatorial plane of the isotropic chart the optical geometry is
\begin{equation}
h=e^{\mu-\nu}\left(dr^2+r^2d\varphi^2\right)=\varrho^2\left(dX^2+d\varphi^2\right),
\label{eq:optical}
\end{equation}
where $\varrho=Re^{-\nu/2}=\Vopt^{-1/2}$ is its circumferential radius. Its meridians have length $dm=\varrho\,dX$, and $d\varrho/dm=-D/2$, with $D=V-U-2$ as in the proof of Proposition~\ref{prop:rings}. Where $|D|<2$, which holds on every orbit to which the proposition applies, (\ref{eq:optical}) is the metric of a surface of revolution in Euclidean space (Fig.~\ref{fig:optical}). A parallel of such a surface is a geodesic exactly where $d\varrho/dm=0$, and every geodesic obeys Clairaut's relation $\varrho\sin\psi=b$, where $\psi$ is its angle to the meridians. For a null geodesic $b=L/E$ is the impact parameter, and the optical length along it is the elapsed Killing time $t$. Null geodesics are therefore confined to the region $\varrho\geq b$, so that the parallel at a neck, where $\varrho$ is minimal, is an unstable light ring and the parallel at a bulge is a stable one. Abramowicz et al.~\cite{Abramowicz1997} used this picture for the trapped modes of ultracompact stars, whose optical geometry has a neck with a bulge inside it, and the necks of~\cite{KRS2000,KRS2001} are of the same kind.

At a matter-bearing equilibrium on the line~(\ref{eq:eos-line}), $D=(k-1)(U+2)$ is a nonzero constant unless $k=1$, so the optical geometry of each self-similar solution is a cone, $\varrho\propto m$, without light rings. At $\gamma=\tfrac13$ the cone degenerates into a cylinder. Every parallel is then a light ring, and the rescalings $R\to\lambda R$ of the self-similar solution act as translations along the cylinder (Fig.~\ref{fig:optical}a). On a periodic core $\varrho$ is a periodic function of $m$ by~(\ref{eq:Vopt-P}), and the cylinder becomes a periodic tube with one neck and one bulge per period (Fig.~\ref{fig:optical}b). Translation by one period is the discrete rescaling $R\to e^{\Delta\ln R}R$ of the core, which multiplies the metric by a constant and leaves $h$ unchanged, so all necks are congruent, as are all bulges, and the singular centre lies at infinite optical distance. A null geodesic with $\varrho_{\rm neck}<b<\varrho_{\rm bulge}$ cannot pass a neck and remains in the band about a single bulge bounded by the two parallels on which $\varrho=b$. In the equatorial plane this band is an annulus about a stable light ring, and the infinitely many such annuli are nested towards the centre in the progressions of Proposition~\ref{prop:rings} (Fig.~\ref{fig:optical}c).

The result places these singular cores among ultracompact configurations with light rings. In general relativity the role of radiation is played by stiff matter, whose self-similar sphere has constant optical potential and hence a cylindrical optical geometry, and Karlovini, Rosquist and Samuelsson found configurations with arbitrarily many necks, with a central role played by the equation of state $p=\rho-\rho_s$~\cite{KRS2000,KRS2001}. There the self-similar solution is a focus rather than a centre, so the oscillations about it are not periodic. Here the integrability of the radiation core makes them exactly periodic in $\ln R$: the necks and bulges repeat without bound, congruent in the optical geometry and at geometrically decreasing areal radii, with a ratio fixed by the amplitude of the core. The stable light rings trap null geodesics in the core. In models of smooth ultracompact objects, stable light rings have been associated with long-lived modes and possible nonlinear instabilities~\cite{Cardoso2014,Keir2016}. For the singular interiors considered here, the optical wells provide a setting in which to investigate long-lived wave modes once the perturbation problem and its boundary conditions are specified.

\section{The MIT bag model}
\label{sec:bag}

\begin{figure*}[t]
\centering
\includegraphics[width=\textwidth]{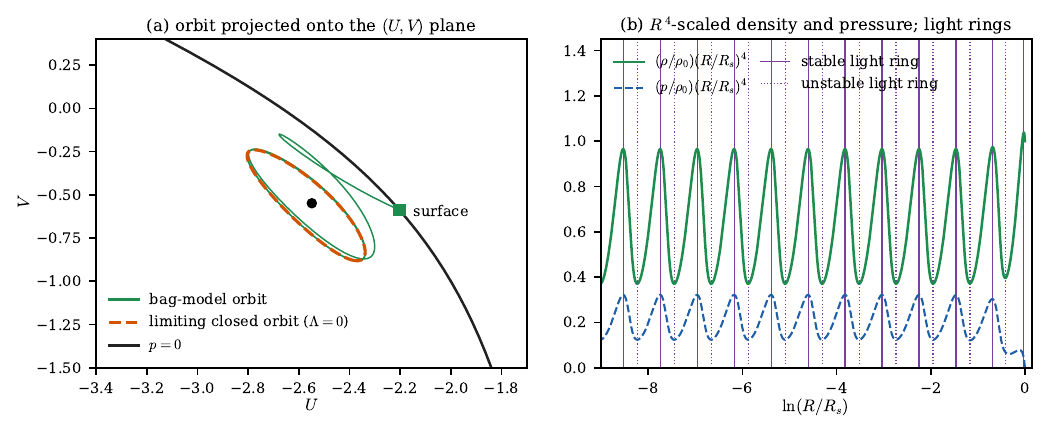}
\caption{An MIT bag configuration ($\gamma=\tfrac13$), started at $P_1+(0.1,0)$ with $\Lambda=10^{-4}$, for which $\Lambda_s\approx144$. (a) The orbit projected onto the $(U,V)$ plane, winding onto a closed orbit of the radiation system in the core and ending on the zero-pressure curve at the surface. (b) Density and pressure in units of $\rho_0=4B$, scaled by $(R/R_s)^4$ to remove the self-similar growth, against $\ln(R/R_s)$; vertical lines mark the stable (solid) and unstable (dotted) light rings}
\label{fig:bag}
\end{figure*}

For $\rho_0\neq0$ the bag constant introduces a length scale, and the variable $\Lambda$ of~(\ref{eq:Lambda}) measures $R^4$ in units of $\epsilon/(8\kappa\rho_0)$; it is positive for $\epsilon>0$ and $\rho_0>0$. An equilibrium with $\Lambda>0$ would need $U=-2$. There $\Phi=4(V-1)(V^2+4)$ vanishes only at $V=1$, which lies on the vacuum line $L_2$ where the system is singular. All equilibria therefore lie in the invariant plane $\Lambda=0$, and the eigenvalue transverse to the plane at an equilibrium $(U_{\rm e},V_{\rm e})$ is $2(U_{\rm e}+2)$. At $P_1$ this lies between $-\tfrac43$ and about $-1.04$ for $\gamma\in(0,1]$, so the plane attracts as $X$ increases, that is, towards the centre.

The MIT bag model sits exactly at the integrable value $\gamma=\tfrac13$. With $\Lambda\neq0$, $H$ is no longer conserved, but its rate of change is simply
\begin{equation}
\dot H=\frac{\Lambda\,Q(U,V)}{2U^4(U+2V)^4(2U+V)^5},
\label{eq:Hdrift}
\end{equation}
so that $H$ freezes as $\Lambda\propto R^4\to0$. For a core trajectory that remains in a compact part of the period annulus, away from the curves $U(U+2V)(2U+V)=0$ on which~(\ref{eq:Hdrift}) is singular, the integral of $|\dot H|$ converges, $H$ tends to a limit, and the core approaches the closed orbit of the radiation system on that level set; the density then grows as $R^{-4}$ times a log-periodic factor. Since the light-ring condition~(\ref{eq:LR-line}) is purely metric, such a core inherits the nested light rings of its limiting orbit, with the spacing of Proposition~\ref{prop:rings} approached as $R\to0$, and its optical geometry approaches the periodic tube of that orbit. This is a local statement; whether bag-model cores generically remain in the annulus requires further investigation. By Proposition~\ref{prop:centre} these singular cores cannot be replaced by regular ones.

Integrating outwards, $\Lambda$ grows and the orbit leaves the plane. A surface $p=0$, where $\rho=\rho_0=4B$, is reached when the orbit meets the curve $3U^2+6UV+8U+8V=0$ at some value $\Lambda_s$. The radius of the configuration is then
\begin{equation}
R_s=\left(\frac{\epsilon\Lambda_s}{32\kappa B}\right)^{1/4}.
\end{equation}
We integrated orbits starting at $P_1+(s,0)$, with $|s|\le0.2$ and $\Lambda$ between $10^{-6}$ and $1$, towards decreasing $X$ for the envelope and towards increasing $X$ for the core. For $-0.2\le s\le0.125$ every such orbit reaches the surface with a monotonic areal radius and $\rho\ge\rho_0$ throughout, with $\Lambda_s$ between about $50$ and $180$. For larger positive $s$ some orbits, and for $s\ge0.2$ all of them, meet $U=-2$ first, which is a turning point of $R$, and are excluded. In the configuration of Fig.~\ref{fig:bag}, with $s=0.1$ and $\Lambda_s\approx144$, the outermost interior light ring is a stable one at $R\approx0.967R_s$, and the ratio between successive rings of each type settles within a few oscillations to $2.193$, the value for the limiting closed orbit. Matching these configurations to an exterior vacuum requires the junction conditions of the theory and is left for future work; until then they are singular perfect-fluid configurations rather than complete stellar models.

\section{Conclusion}
\label{sec:conclusion}

In the pure quadratic $f(G,B)$ model every regular static metric has vanishing central density and pressure, so no affine equation of state $p=\gamma(\rho-\rho_0)$ with $\gamma>0$ admits a regular non-vacuum configuration. The non-vacuum configurations studied here therefore have singular centres, and we have studied their structure with a dynamical systems formulation in scale-invariant variables.

The isotropy condition is there a single autonomous relation for two functions, and its closure matters. Among closures that keep the self-similar solutions as equilibria, the transverse stability of the two vacuum lines is fixed by the field equations, but that of every regular point of the matter-bearing hyperbola other than one unphysical point can be assigned any value. This qualifies the statement in~\cite{HBB} that the sign of the transverse eigenvalue is independent of the closure: it holds on the vacuum lines only, and the two closures used there disagree on an arc of the hyperbola. Agreement between particular closures, such as the minimum norm and gradient normal ones, can reflect an algebraic relation between them rather than a property of the equations. The vacuum sector contains, besides the two lines, a point with $G\neq0$ whose geometry coincides with a power-law vacuum of $f(T)\propto T^2$ gravity~\cite{Golovnev2021}.

The physically meaningful closure is an equation of state, and it gives definite answers. For $p=\gamma\rho$ the closed system has a focus $P_1$, which changes stability exactly at the radiation value $\gamma=\tfrac13$, and a saddle $P_2$, which exchanges stability with the vacuum point in a transcritical bifurcation at $\gamma=\tfrac23$, where its density changes sign. At radiation the system has the first integral~(\ref{eq:H}), built from the Tolman integral and the Noether charge of global scale invariance, and $P_1$ is a nonlinear centre with a family of log-periodic cores.

The main result concerns light. At $\gamma=\tfrac13$, and only there, the line containing the matter-bearing equilibria coincides with the line $V=U+2$ of circular null orbits, so the self-similar radiation solution has a light ring at every radius. Every periodic radiation core with monotonic areal radius splits this degeneracy into exactly one stable and one unstable light ring per period, so that the stable and unstable radii form two interlaced geometric progressions towards the centre, with a common ratio that tends to $2.165$ for small oscillations and grows with amplitude. The optical geometry shows the mechanism directly. Self-similar solutions have conical optical geometries without light rings, except at radiation, where the cone becomes a cylinder. A periodic core turns the cylinder into a periodic tube of congruent necks and bulges at geometrically decreasing areal radii, and null geodesics with impact parameters between the neck and bulge radii are trapped about the bulges. MIT bag-model configurations whose cores approach these orbits carry the same nested light rings, as the example with a surface in Section~\ref{sec:bag} shows. Natural next steps are the junction conditions at the surface, the perturbative stability of configurations with stable light rings, and the model $f=\alpha_0G+\epsilon G^2$, in which a regular centre is no longer excluded while singular cores, where $G$ diverges and the quadratic term dominates, should retain the structure found here.

\medskip
\noindent\textbf{Data Availability Statement} This manuscript has no associated data. [Authors' comment: This is a theoretical study; all results follow from the expressions given in the text.]

\medskip
\noindent\textbf{Code Availability Statement} Code/software will be made available on reasonable request. [Authors' comment: The Mathematica notebook and Python scripts that implement the closures, reproduce the tables and generate the figures are available from the corresponding author.]

\appendix

\section{Explicit closures}
\label{app:gauges}

\begin{figure*}[!tp]
\centering
\includegraphics[width=\textwidth]{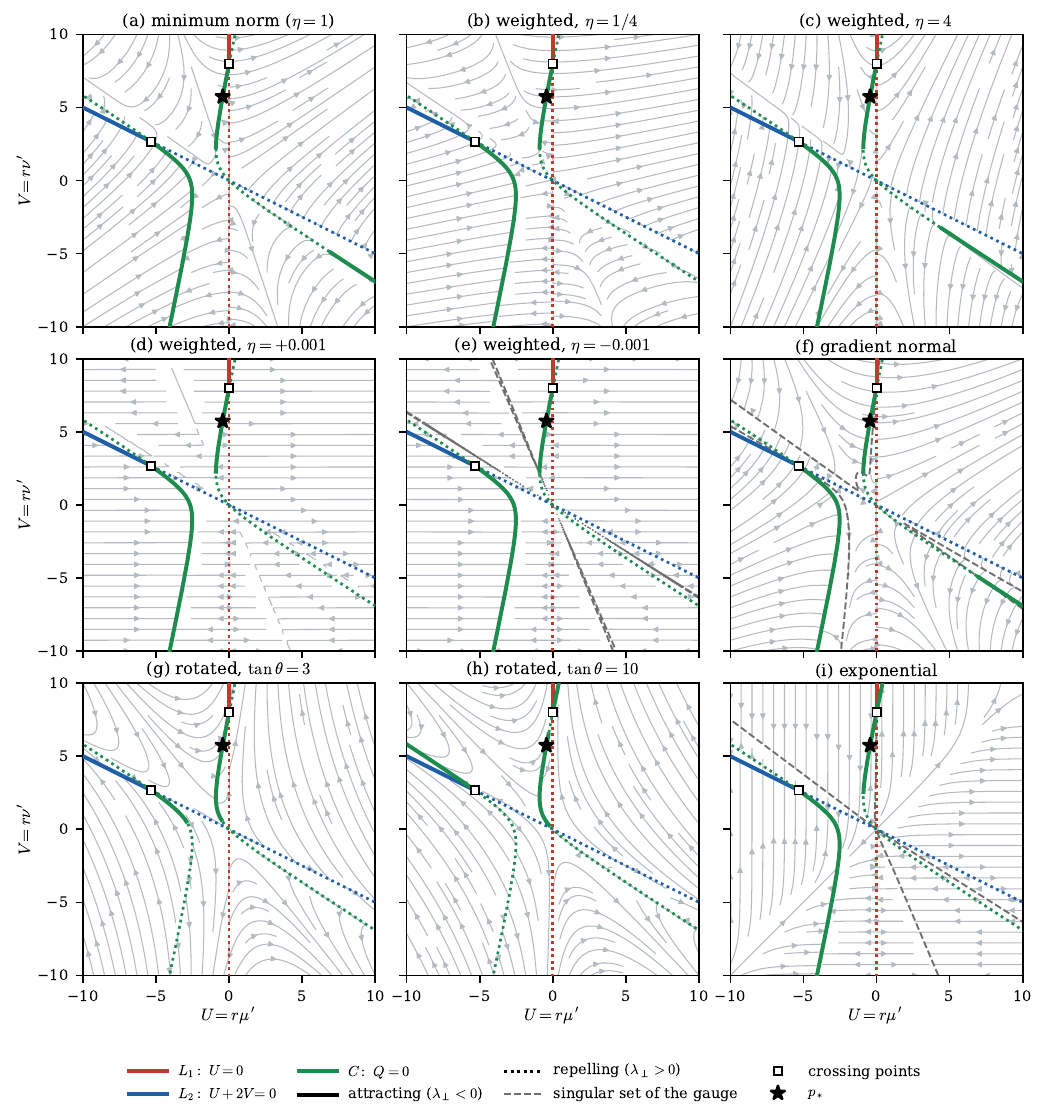}
\caption{Phase portraits for the explicit gauges: (a) the minimum norm gauge, which is the weighted gauge with $\eta=1$; (b)--(e) the weighted gauge with $\eta=\tfrac14$, $4$, $+0.001$ and $-0.001$; (f) the gradient normal gauge; (g), (h) the rotated normal gauge with $\tan\theta=3$ and $10$; (i) the exponential gauge. Line styles as in Fig.~\ref{fig:pair}: each equilibrium curve is solid where it is transversely attracting and dotted where it is repelling. The classification of $L_1$ and $L_2$ is common to all panels; that of $C$ is not. Dashed grey curves are the singular sets of the gauges}
\label{fig:portraits}
\end{figure*}

This appendix collects five explicit admissible gauges, which illustrate Theorem~\ref{thm:main}. Each is specified by its direction field $\bw$, and its function $\Omega$ follows from~(\ref{eq:direction-gauge}). Table~\ref{tab:transitions} lists the points of $C$ at which $\lambda_\perp$ changes sign; between consecutive listed points the sign is constant and was confirmed by direct sampling and by finite-difference evaluation of the Jacobian trace of the full vector field. Fig.~\ref{fig:portraits} shows the phase portraits.

On $L_1$ and $L_2$ the constraint itself permits motion along the line: on $U=0$ it reduces to $4V^2\dot U=0$, leaving $\dot V$ free, and on $U+2V=0$ it reduces to $U^2(\dot U+2\dot V)=0$. These motions are the two vacuum families of Section~\ref{sec:physics}. Admissible gauges freeze them, since every point of $\Phi=0$ becomes an equilibrium, so only their self-similar members are retained. Along a general member of the first family, with $V=V(X)$, linearising the constraint itself gives $\delta\dot U=[(8-V)/2-2\dot V/V]\,\delta U$, and along a general member of the second, with $s=U+2V$, $\delta\dot s=[(3U+16)/4-2\dot U/U]\,\delta s$. Neither rate involves a closure; the eigenvalues~(\ref{eq:L12}) are the special cases $\dot V=0$ and $\dot U=0$.

\paragraph{Minimum norm and weighted norm gauges.}
Minimising $\dot U^2+\eta^{-1}\dot V^2$ ($\eta>0$) under the constraint gives $\bw=(\alpha,\eta\beta)$:
\begin{align}
\dot U&=\frac{\alpha\Phi}{\alpha^2+\eta\beta^2},\qquad \dot V=\frac{\eta\beta\Phi}{\alpha^2+\eta\beta^2},\nonumber\\
\Omega&=\frac{(\eta-1)\alpha\beta}{(\alpha^2+\eta\beta^2)|\bn|^2}.
\label{eq:weighted}
\end{align}
For $\eta=1$ this is the minimum norm gauge of~\cite{HBB}, which chooses the admissible velocity of least Euclidean length, since $|\dot\by|^2=\Phi^2/|\bn|^2+h^2|\bn|^2$; then $\bw=\bn$, $\Omega=0$ and $\lambda_\perp=\lmin$. On $C$, $\lmin$ vanishes, apart from the crossing points and the origin, only at
\begin{align}
a_1&\approx(6.862707,\,-4.866226),\nonumber\\
a_2&\approx(-0.906062,\ 2.154013),
\label{eq:a12}
\end{align}
both on $C_+$, the real roots on $C$ of the resultant $-96V^2(147V^4+280V^3-1472V^2+2304V-4096)$. The minimum norm classification is: on $C_+$, attracting for $V<-4.8662$, repelling for $-4.8662<V<0$ and $0<V<2.1540$, attracting for $2.1540<V<8$ and repelling for $V>8$; on $C_-$, attracting for $V<\tfrac83$ and repelling for $V>\tfrac83$.

For general $\eta$ the transverse eigenvalue is a weighted average,
\begin{align}
\lambda_\perp^{(\eta)}&=\frac{\alpha\Phi_U+\eta\beta\Phi_V}{\alpha^2+\eta\beta^2}\nonumber\\
&=\frac{\alpha^2}{\alpha^2+\eta\beta^2}\,\frac{\Phi_U}{\alpha}+\frac{\eta\beta^2}{\alpha^2+\eta\beta^2}\,\frac{\Phi_V}{\beta},
\label{eq:weighted-average}
\end{align}
so its sign is independent of $\eta$ exactly where $\Phi_U/\alpha$ and $\Phi_V/\beta$ have the same sign. These agree on $L_1$ and $L_2$, but not on large parts of $C$. At the point of $C_+$ with $V=-4$ ($U\approx5.5497$), for example, $\lambda_\perp^{(1/4)}\approx15.87$, $\lambda_\perp^{(1)}\approx6.39$ and $\lambda_\perp^{(4)}\approx-17.25$. As $\eta$ increases from $0$ to $\infty$, the left transition point on $C_+$ moves from $V=-\infty$ to $V=-\tfrac83$. The limit $\eta\to0$ is the closure $\dot V=0$ discussed in Section~\ref{sec:closure}. It is singular where $\alpha=0$, which on $C$ happens at $V=8(2-\sqrt3)\approx2.144$ on $C_+$ and at $V=8(2+\sqrt3)\approx29.86$ on $C_-$; beyond the latter point $C_-$ is attracting rather than repelling.

The formulae~(\ref{eq:weighted}) remain an admissible gauge for $\eta<0$, although the quadratic form is then indefinite and the velocity is a stationary point rather than a minimum. The gauge acquires the singular set $\alpha^2+\eta\beta^2=0$, which is four lines through the origin, $\alpha=\pm\sqrt{-\eta}\,\beta$. As $\eta\to0^-$ they close in pairwise on the two lines $\alpha=0$, namely $V/U=(-3\pm\sqrt3)/2$; for $\eta=-0.001$ their slopes are $V/U\approx-0.6296,\,-0.6381$ and $-2.3072,\,-2.4252$. Inside the thin wedges between each pair the field is reversed and unbounded, so the phase portrait for small negative $\eta$ looks radically different from that for small positive $\eta$ (Fig.~\ref{fig:portraits}d,e). The invariant content is untouched: on $L_1$ and $L_2$, which none of the singular lines meet for $-\tfrac14<\eta<0$, the eigenvalues are~(\ref{eq:L12}). On $C$ the classification acquires additional sign changes through poles, at $V\approx2.088$ and $2.197$ on $C_+$ and at $V\approx21.88$ and $45.53$ on $C_-$ for $\eta=-0.001$. At $\eta=-\tfrac14$ the line $L_2$ itself becomes singular, since $\alpha^2+\eta\beta^2=(1+4\eta)U^4$ there.

\paragraph{Gradient normal gauge.}
Taking the velocity normal to the level sets of $\Phi$, $\bw=\nabla\Phi$, gives
\begin{equation}
\dot\by=\frac{\Phi}{D}\,\nabla\Phi,\qquad \Omega=\frac{W}{D\,|\bn|^2},\qquad
\lambda_\perp^{(\nabla)}=\frac{|\nabla\Phi|^2}{D}.
\label{eq:gradient}
\end{equation}
Since $\lambda_\perp^{(\nabla)}\,\lmin=|\nabla\Phi|^2/|\bn|^2>0$ at regular points, this gauge always reproduces the minimum norm classification. This is not evidence of gauge independence: the two gauges are linked by the identity $D^2+W^2=|\bn|^2|\nabla\Phi|^2$, and the transition points $a_1$, $a_2$ of~(\ref{eq:a12}) become poles of $\lambda_\perp^{(\nabla)}$, where the gauge is singular. Eliminating $V$ between $Q=0$ and $D=0$ gives $32U^5(3U+16)(441U^4-952U^3-11072U^2-20224U-10240)$, whose real roots reproduce $a_1$, $a_2$ and the crossing points.

\paragraph{Rotated normal gauge.}
Rotating the minimum norm direction through a constant angle $\theta$, $|\theta|<\pi/2$, gives $\bw=\cos\theta\,\bn+\sin\theta\,\bnp$, $\Omega=\tan\theta/|\bn|^2$ and
\begin{equation}
\lambda_\perp^{(\theta)}=\lmin+\tan\theta\,\frac{W}{|\bn|^2}.
\label{eq:rotated-lambda}
\end{equation}
The correction is linear in $\tan\theta$, with a coefficient that vanishes identically on $L_1$ and $L_2$ but not on $C$. At $\tan\theta=3$ two transition points appear on $C_-$, at $V\approx0.417$ and $V\approx6.078$. At $\tan\theta=10$ the classification of $C_-$ is reversed on both of its arcs: $C_-$ is repelling for $V<\tfrac83$ and attracting for $V>\tfrac83$ (verified for $|V|\le200$). At the point $(1,\,6+\sqrt{47})$ of $C_+$ one has $\lmin\approx2.0329$ and $W/|\bn|^2\approx-0.7153$, so this point, repelling in the minimum norm gauge, is attracting for $\tan\theta>2.842$.

\paragraph{Exponential gauge.}
As an example with no geometric motivation, take $\bw=(e^U,e^V)$, for which
\begin{equation}
\lambda_\perp^{(\exp)}=\frac{e^U\Phi_U+e^V\Phi_V}{\alpha e^U+\beta e^V}.
\label{eq:exponential}
\end{equation}
It is singular on the curve $\alpha e^U+\beta e^V=0$, which meets $C_+$ at $V\approx7.9933$, just below the crossing point $(0,8)$. On $C$ the numerator vanishes only at $V\approx2.418$ on $C_+$. The resulting classification differs from the minimum norm one on the arc $V<-4.8662$ of $C_+$, which is repelling here, and on the arc $V>8$, which is attracting here; at $V=-6$, $\lambda_\perp^{(\exp)}\approx32.86$. The classification of $C_-$ coincides with the minimum norm one for $|V|\le40$.

\begin{table*}[t]
\centering
\caption{Points of the hyperbola at which the transverse eigenvalue changes sign, located by root finding along $C_\pm$ for $|V|\le60$ ($|V|\le200$ for the rotated gauges, $|V|\le40$ for the exponential gauge). $^\dagger$The origin, excluded in every gauge. $^\ddagger$Crossing points with $L_1$ or $L_2$. On $L_1$ and $L_2$ all gauges give~(\ref{eq:L12}), and every gauge gives $\lambda_\perp=-\tfrac43$ at $p_*$ ($V_*\approx5.7379$)}
\label{tab:transitions}
\footnotesize
\setlength{\tabcolsep}{3.5pt}
\begin{tabular}{lll}
\toprule
Gauge & Sign changes of $\lambda_\perp$ on $C_+$ (values of $V$) & Sign changes on $C_-$ \\
\midrule
Minimum norm ($\eta=1$) & $-4.8662$,\ $0^\dagger$,\ $2.1540$,\ $8^\ddagger$ & $8/3^\ddagger$ \\
Weighted, $\eta=1/4$ & $-13.5613$,\ $0^\dagger$,\ $2.1468$,\ $8^\ddagger$ & $8/3^\ddagger$ \\
Weighted, $\eta=4$ & $-3.1973$,\ $0^\dagger$,\ $2.1679$,\ $8^\ddagger$ & $8/3^\ddagger$ \\
Weighted, $\eta\to0$ ($\dot V=0$) & $0^\dagger$,\ pole $2.1436$,\ $8^\ddagger$ & $8/3^\ddagger$,\ pole $29.856$ \\
Weighted, $\eta=-0.001$ & $0^\dagger$,\ pole $2.088$,\ $2.143$,\ pole $2.197$,\ $8^\ddagger$ & $8/3^\ddagger$,\ pole $21.88$,\ $29.35$,\ pole $45.53$ \\
Gradient normal & poles at $-4.8662$,\ $2.1540$; otherwise as minimum norm & $8/3^\ddagger$ \\
Rotated, $\tan\theta=-1$ & $-2.1088$,\ $0^\dagger$,\ $3.2517$,\ $8^\ddagger$ & $8/3^\ddagger$ \\
Rotated, $\tan\theta=1/2$ & $-9.0786$,\ $0^\dagger$,\ $1.5502$,\ $8^\ddagger$ & $8/3^\ddagger$ \\
Rotated, $\tan\theta=3$ & $0^\dagger$,\ $0.5394$,\ $8^\ddagger$,\ $11.8863$ & $0.4171$,\ $8/3^\ddagger$,\ $6.0778$ \\
Rotated, $\tan\theta=10$ & $0^\dagger$,\ $0.1876$,\ $6.6151$,\ $8^\ddagger$ & $8/3^\ddagger$ (reversed) \\
Exponential & $0^\dagger$,\ $2.418$,\ pole at $7.9933$,\ $8^\ddagger$ & $8/3^\ddagger$ \\
\bottomrule
\end{tabular}
\end{table*}

\section{Behaviour at infinity}
\label{app:infinity}

Since $U$ and $V$ are unbounded, the global phase portrait requires a compactification~\cite{Perko}. The function $\Phi$ has degree four, with quartic part $\Phi_4=U(U+2V)(3U^2+4UV-V^2)$, and $\alpha$ and $\beta$ are homogeneous quadratics. Consider a gauge whose direction field has a homogeneous leading part $\tilde\bw$ with $\tilde\bw\cdot\bn\not\equiv0$. Its vector field then has leading homogeneous part
\begin{equation}
(F_2,G_2)=\frac{\Phi_4}{\tilde\bw\cdot\bn}\,\tilde\bw
\end{equation}
of degree two, whatever the degree of $\tilde\bw$. This covers the minimum norm, weighted, gradient normal and rotated gauges, but not a general $\Omega$: a nonzero constant $\Omega$, for instance, has leading part $\Omega\bnp$ with $\bnp\cdot\bn=0$, and its vector field has leading degree six. In the chart $u=V/U$, $z=1/U$ the reparametrisation $dX=z\,d\tau$ gives
\begin{align}
\frac{du}{d\tau}&=G_2(1,u)-u\,F_2(1,u)+O(z),\nonumber\\
\frac{dz}{d\tau}&=-z\,F_2(1,u)+O(z^2),
\label{eq:chart}
\end{align}
and analogously in the chart $v=U/V$, $z=1/V$ for the direction $U=0$. Equilibria at infinity are the zeros of the angular function $h(u)=G_2(1,u)-uF_2(1,u)$, with normal eigenvalue $-F_2(1,u)$ and angular eigenvalue $h'(u)$. Because the leading degree is even, $(F_2,G_2)$ is invariant under $(U,V)\mapsto(-U,-V)$. This map therefore carries orbits near a point at infinity to orbits near the antipodal point with the direction of $X$ reversed; equivalently, $z<0$ on the ray $U<0$ and $dX=z\,d\tau$ reverses orientation there. Antipodal equilibria have eigenvalues of opposite sign: stable and unstable nodes are interchanged, and saddles remain saddles. Below, eigenvalues and types refer to the ray $U>0$, and to $V>0$ for the direction $U=0$.

For such a gauge $h=\Phi_4\,(\tilde w_2-u\,\tilde w_1)/(\tilde\bw\cdot\bn)$ at $U=1$, and the equilibria at infinity split into two classes. The first consists of the zeros of $\Phi_4$, namely the asymptotic directions $U=0$, $V/U=-\tfrac12$ and $V/U=2\pm\sqrt7$ of $L_1$, $L_2$ and $C$. These are present in every such gauge and are the endpoints of the equilibrium curves; the normal eigenvalue vanishes there. The two endpoints of $C$ on the ray $U>0$ belong to $C_+$, and those on the ray $U<0$ to $C_-$. The second class consists of the zeros of $\tilde w_2-u\tilde w_1$, which depend on the gauge. At the endpoints of $L_1$ and $L_2$ the angular eigenvalues are $-\tfrac12$ and $\tfrac34$ in every gauge, as follows from $\bn\parallel\nabla\Phi$ on those lines. At the endpoints of $C$ they depend on the gauge (Table~\ref{tab:infinity}): at $V/U=2-\sqrt7$ the value is $-1.717$ in the minimum norm gauge but $+2.366$ in the rotated gauge with $\tan\theta=3$, and at $V/U=2+\sqrt7$ it is $+1.476$ and $-0.918$ respectively. In the minimum norm gauge $h(u)$ factorises as
\begin{equation}
h(u)=\frac{(u+3)(2u+1)(2u^2-1)(u^2-4u-3)}{4\,(2u^4+12u^3+32u^2+30u+9)},
\end{equation}
and in the gradient normal gauge as
\begin{equation}
h(u)=\frac{(2u+1)(u^2-4u-3)\,P_4(u)}{2\,P_5(u)},
\end{equation}
with $P_4=u^4-7u^3-18u^2+u+5$ and $P_5=2u^5-8u^4-57u^3-142u^2-122u-33$.

\begin{figure*}[!tp]
\centering
\includegraphics[width=\textwidth]{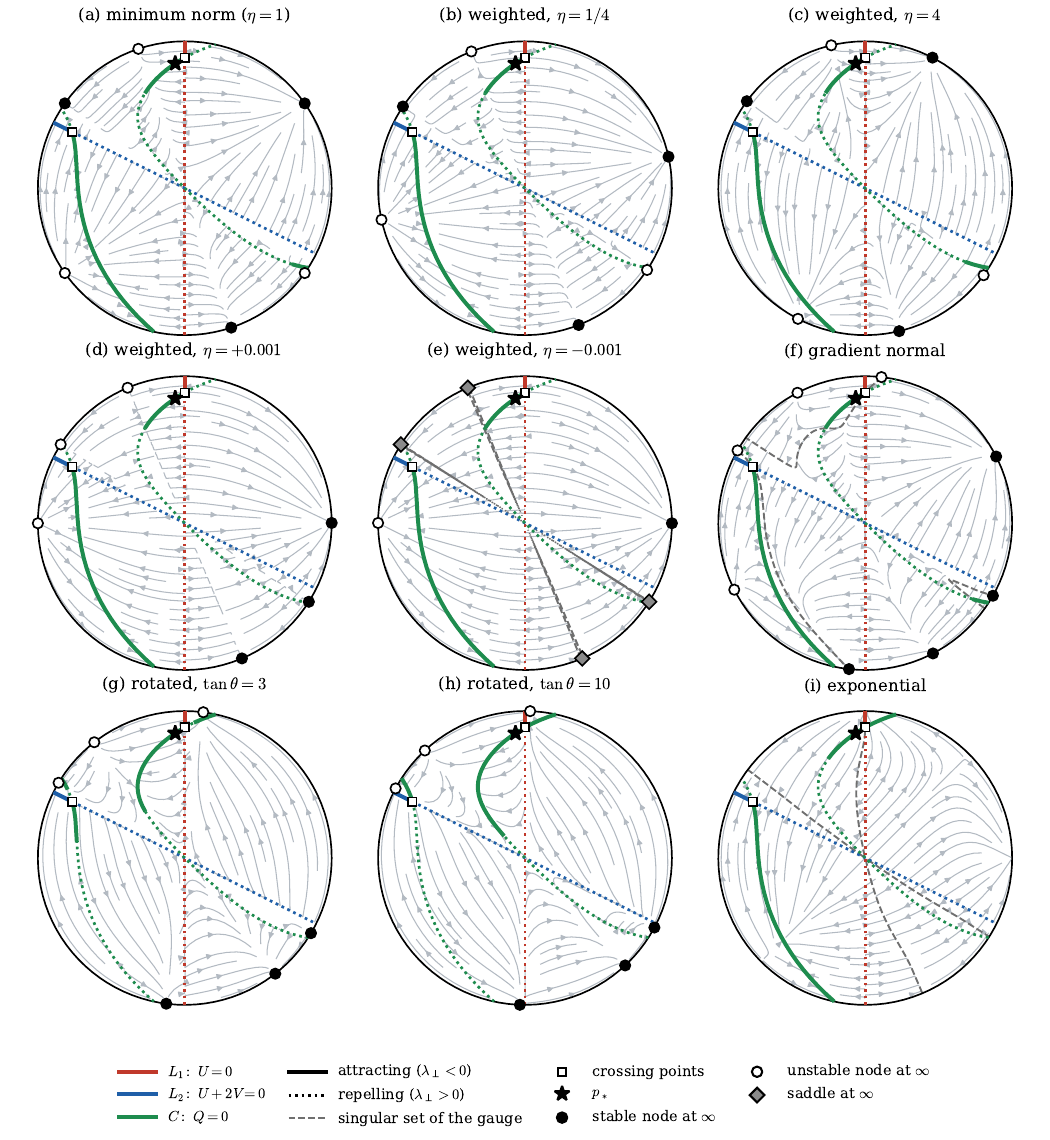}
\caption{Poincar\'e discs for the gauges of Fig.~\ref{fig:portraits}, drawn in the coordinates $(U,V)/(1+\sqrt{U^2+V^2})$, with the same line styles. For the gauges (a)--(h), whose vector fields have homogeneous leading parts, filled and open circles mark stable and unstable nodes at infinity and diamonds mark saddles; antipodal equilibria at infinity have opposite types. The exponential gauge (i) is discussed at the end of \ref{app:infinity}}
\label{fig:discs}
\end{figure*}

The isolated equilibria at infinity (Table~\ref{tab:infinity-iso}) change position, number and type between gauges. For $\eta=+0.001$ all three are stable nodes on the ray $U>0$, and hence unstable nodes on the antipodal ray, while for $\eta=-0.001$ two of them become saddles, lying inside the reversed wedges between the singular lines (Fig.~\ref{fig:discs}d,e). These equilibria, including the attractors that determine the fate of orbits which do not approach the finite equilibrium set, are artefacts of the closure; only the endpoints of $L_1$ and $L_2$ have the same angular eigenvalue in every gauge.

The exponential gauge is not of polynomial-ratio type, and its direction field has no homogeneous leading part. It tends to $(1,0)$ where $U-V\to+\infty$ and to $(0,1)$ where $V-U\to+\infty$, so on the circle at infinity it acts as the closure $\dot V=0$ on one side of the diagonal $U=V$ and as $\dot U=0$ on the other, the limits $\eta\to0$ and $\eta\to\infty$ of the weighted family. The endpoints of $L_1$, $L_2$ and $C$ remain equilibria at infinity. So is the direction $V/U=0$, but on the ray $U>0$ only, so the antipodal symmetry of the other gauges is lost (Fig.~\ref{fig:discs}i). The diagonal $U=V$, where the leading closure switches, is an invariant line, since $\bw\propto(1,1)$ there, but its directions at infinity are not ordinary equilibria. On the ray $U>0$ the angular function tends to $-\tfrac{9}{11}$ as $V/U\to1$ from below and to $\tfrac97$ from above, while it vanishes on the diagonal itself, so the compactified angular field is discontinuous there and the diagonal is an invariant direction rather than an equilibrium that can be classified by linearisation.

\begin{table}[t]
\centering
\caption{Angular eigenvalues at the two endpoints of $C_+$ on the circle at infinity (ray $U>0$); on the antipodal ray, the endpoints of $C_-$, the signs are reversed. At the endpoints of $L_1$ and $L_2$ the angular eigenvalues are $-\tfrac12$ and $\tfrac34$ in every gauge, and the normal eigenvalue vanishes at all endpoints}
\label{tab:infinity}
\small
\begin{tabular}{lcc}
\toprule
Gauge & $V/U=2-\sqrt7$ & $V/U=2+\sqrt7$\\
\midrule
Minimum norm & $-1.717$ & $1.476$\\
Weighted, $\eta=1/4$ & $-1.335$ & $1.647$\\
Weighted, $\eta=4$ & $-1.816$ & $0.955$\\
Weighted, $\eta=+0.001$ & $10.94$ & $1.708$\\
Weighted, $\eta=-0.001$ & $13.97$ & $1.709$\\
Gradient normal & $-2.796$ & $1.907$\\
Rotated, $\tan\theta=3$ & $2.366$ & $-0.918$\\
Rotated, $\tan\theta=10$ & $11.89$ & $-6.504$\\
\bottomrule
\end{tabular}
\end{table}

\begin{table*}[t]
\centering
\caption{Isolated equilibria at infinity, the zeros of $\tilde w_2-u\tilde w_1$, with eigenvalues and type on the ray $U>0$. On the antipodal ray the eigenvalues change sign, so stable and unstable nodes are interchanged and saddles remain saddles. A negative normal eigenvalue means that orbits reach infinity as $r$ increases. The gradient normal gauge is in addition singular in the directions $V/U\approx-0.7158$, $-0.5705$ and $8.4629$, where the leading part of $D$ vanishes, and the weighted gauge with $\eta=-0.001$ along its four singular lines}
\label{tab:infinity-iso}
\small
\begin{tabular}{lrrrl}
\toprule
Gauge & Direction $V/U$ & Normal eigenvalue & Angular eigenvalue & Type ($U>0$)\\
\midrule
Minimum norm & $-3$ & $-1.500$ & $-8.500$ & stable node\\
 & $-1/\sqrt2$ & $0.187$ & $4.995$ & unstable node\\
 & $1/\sqrt2$ & $-0.520$ & $-0.662$ & stable node\\
\midrule
Weighted, $\eta=1/4$ & $-2.5502$ & $-1.473$ & $-21.75$ & stable node\\
 & $-0.6694$ & $0.064$ & $1.764$ & unstable node\\
 & $0.2197$ & $-0.523$ & $-0.584$ & stable node\\
\midrule
Weighted, $\eta=4$ & $-4.2062$ & $-1.638$ & $-5.318$ & stable node\\
 & $-0.7349$ & $0.302$ & $17.01$ & unstable node\\
 & $1.9411$ & $-0.397$ & $-0.588$ & stable node\\
\midrule
Weighted, $\eta=+0.001$ & $-2.3668$ & $-1.472$ & $-4419$ & stable node\\
 & $-0.6342$ & $-0.028$ & $-83.83$ & stable node\\
 & $0.0010$ & $-0.500$ & $-0.501$ & stable node\\
\midrule
Weighted, $\eta=-0.001$ & $-2.3652$ & $-1.472$ & $4411$ & saddle\\
 & $-0.6338$ & $-0.029$ & $86.03$ & saddle\\
 & $-0.0010$ & $-0.500$ & $-0.500$ & stable node\\
\midrule
Gradient normal & $-1.9239$ & $-1.515$ & $-5.354$ & stable node\\
 & $-0.5688$ & $-0.155$ & $-25.92$ & stable node\\
 & $0.5085$ & $-0.528$ & $-0.766$ & stable node\\
 & $8.9842$ & $0.697$ & $31.76$ & unstable node\\
\midrule
Rotated, $\tan\theta=3$ & $-1.2788$ & $-2.242$ & $-5.671$ & stable node\\
 & $-0.5958$ & $-0.108$ & $-0.728$ & stable node\\
 & $7.8746$ & $0.517$ & $1.132$ & unstable node\\
\midrule
Rotated, $\tan\theta=10$ & $-1.0730$ & $-5.577$ & $-15.05$ & stable node\\
 & $-0.5375$ & $-0.202$ & $-0.775$ & stable node\\
 & $28.6105$ & $3.946$ & $10.71$ & unstable node\\
\bottomrule
\end{tabular}
\end{table*}

\section{The same mechanism in general relativity}
\label{app:GR}

In general relativity the isotropy condition~(\ref{eq:GRconstraint}) has the form $\dot U+\dot V=\Phi_{\rm GR}$, with $\Phi_{\rm GR}=2(U+V)+\tfrac12(U^2+2UV-V^2)$ and constant normal $\bn=(1,1)$. The equilibrium set of every admissible gauge is the hyperbola $\Phi_{\rm GR}=0$, whose points are the self-similar perfect fluid solutions in isotropic coordinates, and Proposition~\ref{prop:lin} applies unchanged. Here $W=\Phi_{{\rm GR},V}-\Phi_{{\rm GR},U}=-2V$, which vanishes on the hyperbola only at $(0,0)$ and $(-4,0)$, with $\lmin=2+U$ equal to $2$ and $-2$ respectively. Both points are flat space: the first trivially, the second because $e^{\mu}\propto r^{-4}$ is flat space under the inversion $r\mapsto1/r$. Their eigenvalues differ in sign only through orientation: $d\ln R/dX=(U+2)/2$ equals $1$ at the first point and $-1$ at the second, so both have eigenvalue $2$ per unit $\ln R$. Consequently, in general relativity the transverse stability of every non-flat self-similar perfect fluid in this scheme is gauge dependent. A concrete example is the singular stiff-fluid sphere $(U,V)=(-2+\sqrt2,\sqrt2)$, the self-similar solution with $p=\rho=1/(2\kappa R^2)$. There $\lmin=\sqrt2$ and $W=-2\sqrt2$, so in the gauge with constant $\Omega$
\begin{equation}
\lambda_\perp=\sqrt2\,(1-2\Omega).
\label{eq:stiff}
\end{equation}
The same exact solution is transversely repelling for $\Omega=0$ and attracting for $\Omega=1$; since $U+2=\sqrt2>0$, this holds with respect to the areal radius as well.

The same structure appears in the more familiar Buchdahl form. With $e^{\nu}=y^2(x)$, $e^{-\lambda}=Z(x)$ and $x=Cr^2$, the isotropy condition in Schwarzschild coordinates reads
\begin{multline}
4x^2Z\,\frac{d^2y}{dx^2}+2x^2\frac{dZ}{dx}\frac{dy}{dx}\\
+\left(x\frac{dZ}{dx}-Z+1\right)y=0 .
\label{eq:buchdahl}
\end{multline}
Every term is invariant under $x\mapsto kx$, so with $t=\ln x$ and a prime denoting $d/dt$ the equation becomes
\begin{equation}
4Z\left(y''-y'\right)+2Z'y'+\left(Z'-Z+1\right)y=0,
\end{equation}
with no explicit dependence on $t$. Because~(\ref{eq:buchdahl}) is also linear and homogeneous in $y$, the substitution $w=y'/y$ lowers the order and gives
\begin{equation}
(2w+1)\,Z'+4Z\,w'=\Phi_{\rm B},\qquad \Phi_{\rm B}=Z-1-4Zw(w-1).
\label{eq:buchdahl-constraint}
\end{equation}
This is again a single autonomous relation for two unknowns, with normal $\bn=(2w+1,4Z)$ in the $(Z,w)$ plane. Its equilibria, on the curve $Z=1/(1+4w-4w^2)$, are the self-similar solutions with $e^{-\lambda}$ constant and $e^{\nu}\propto x^{2w}$. Writing $w=(1-n)/2$ gives $Z=1/(2-n^2)$, the single-power member of Tolman's solution~VI~\cite{Tolman1939}. The two formulations are related by
\begin{equation}
Z=\frac{(U+2)^2}{4},\qquad w=\frac{V}{2(U+2)},\qquad \frac{d\ln x}{dX}=U+2,
\end{equation}
under which $\Phi_{\rm B}=\Phi_{\rm GR}/2$. For~(\ref{eq:buchdahl-constraint}) one finds $W=-16Zw$, which vanishes on the equilibrium curve only at $(Z,w)=(1,0)$, Minkowski space. The direction field $\bw=(0,1)$, which imposes $Z'=0$, reduces~(\ref{eq:buchdahl}) to the Euler equation $4Zx^2\,d^2y/dx^2+(1-Z)y=0$, whose orbits are the full two-power Tolman~VI family. A known exact solution is thus the orbit set of one particular gauge, and the classical practice of prescribing one potential and solving the linear equation~(\ref{eq:buchdahl}) for the other is a closure of exactly the kind considered here.


\end{document}